\documentclass[journal,twoside,web]{ieeecolor}
\usepackage{jsen}
\usepackage{amsmath,amssymb,amsfonts}
\usepackage{algorithmic}
\usepackage{graphicx}
\usepackage{textcomp}
\usepackage{hyperref}
\usepackage{wrapfig}
\usepackage{adjustbox}
\usepackage[style=ieee,url=false,doi=false,isbn=false,backend=biber]{biblatex}
\usepackage{placeins}
\usepackage{dsfont}
\usepackage{subfigure}
\usepackage{subcaption}
\usepackage{rotating}

\DeclareMathAlphabet{\mathbbold}{U}{bbold}{m}{n}
\NewDocumentCommand{\rot}{O{90} O{1em} m}{\makebox[#2][l]{\rotatebox{#1}{#3}}}

\def\BibTeX{{\rm B\kern-.05em{\sc i\kern-.025em b}\kern-.08em
    T\kern-.1667em\lower.7ex\hbox{E}\kern-.125emX}}
\definecolor{abstractbg}{rgb}{0.89804,0.94510,0.83137}
\AtEveryBibitem{\clearfield{pages}}
\AtEveryBibitem{\clearfield{volume}}
\AtEveryBibitem{\clearfield{number}}

\begin{document}
\title{A Framework for Hybrid Collective Inference in Distributed Sensor Networks}
\author{Andrew~Nash, Dirk~Pesch, Krishnendu~Guha
\thanks{This publication has emanated from research conducted with the financial support of Taighde Éireann – Research Ireland under Grant number 18/CRT/6222 (ADVANCE CRT). For the purpose of Open Access, the author has applied a CC BY public copyright licence to any Author Accepted Manuscript version arising from this submission.}
\thanks{The authors are with the School of Computer Science and IT, University College Cork, Ireland. E-mail: a.nash@cs.ucc.ie, dirk.pesch@ucc.ie, kguha@ucc.ie }
}

\IEEEtitleabstractindextext{%
\fcolorbox{abstractbg}{abstractbg}{%
\begin{minipage}{\textwidth}%
\begin{wrapfigure}[24]{r}{3in}%
\includegraphics[width=3in]{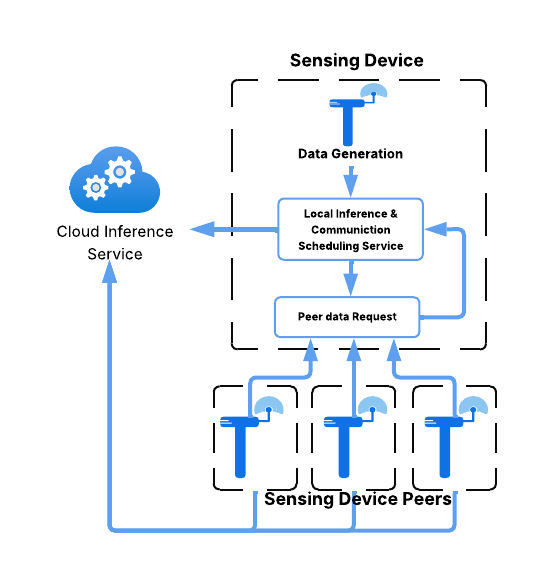}%
\end{wrapfigure}%
\begin{abstract}
With the ever-increasing range of applications of Internet in Things (IoT) and sensor networks, challenges are emerging in various categories of classification tasks. Applications such as vehicular networking, UAV swarm coordination and cyber-physical systems require global classification over distributed sensors, with tight constraints on communication and computation resources. There has been much research in decentralized and distributed data-exchange for communication-efficient collective inference. Likewise, there has been considerable research involving the use of cloud and edge computing paradigms for efficient task allocation. To the best of our knowledge, there has been no research on the integration of these two concepts to create a hybrid cloud and distributed approach that makes dynamic runtime communication strategy decisions. In this paper, we focus on aspects of combining distributed and hierarchical communication and classification approaches for collective inference. We derive optimal policies for agents that implement this hybrid approach, and evaluate their performance under various scenarios of the distribution of underlying data. Our analysis shows that this approach can maintain a high level of classification accuracy (comparable to that of centralised joint inference over all data), at reduced theoretical communication cost. We expect there is potential for our approach to facilitate efficient collective inference for real-world applications, including instances that involves more complex underlying data distributions.
\end{abstract}

\begin{IEEEkeywords}
Cloud, Collective Inference, Communication Efficient, Distributed, Edge, Energy Efficient, Fog, Hybrid, IoT, Multi Agent Systems, Wireless Sensor Networks
\end{IEEEkeywords}
\end{minipage}}}

\maketitle

\section{Introduction}
\label{sec:introduction}

\IEEEPARstart{M}{any} emerging applications, including UAV swarms \cite{yumingDecentralizedConsensusInferencebased2025}, IoT platforms \cite{fagbohungbeEfficientPrivacyPreserving2022}, and Intelligent Transportation systems \cite{shuCollaborativeIntrusionDetection2021} require efficient collaborative data sharing between nodes to achieve overarching collective inference tasks. These often require a very large number of sensing devices which communicate via limited, expensive or otherwise constrained wireless communication links. Many of these devices are also constrained in their computational ability. In the case of both UAVs \cite{lakiotakisJointOptimizationUAV2019} and microcontroller-based low-power IoT sensor nodes \cite{sadlerFundamentalsEnergyconstrainedSensor2005}, the individual devices involved can also be constrained by their limited battery capacities. Various approaches in the literature have been proposed to circumvent these constraints and address the challenge of creating efficient and accurate frameworks for collective inference, in particular:

\begin{enumerate}
    \item The use of cloud and edge computing to offload data to locations where processing is expected to be cheapest \cite{shuCollaborativeIntrusionDetection2021}.
    \item Principles of early-exit computing to split tasks between constrained devices and more powerful edge or cloud computing devices \cite{baccarelliLearningintheFogLiFoDeep2021}.
    \item The use of inter-agent data sharing to form consensus between distributed sensing devices without the use of any centralised computing service \cite{kazariDecentralizedAnomalyDetection2023}.
\end{enumerate}

\begin{figure}
    \centering
    \includegraphics[width=0.8\linewidth]{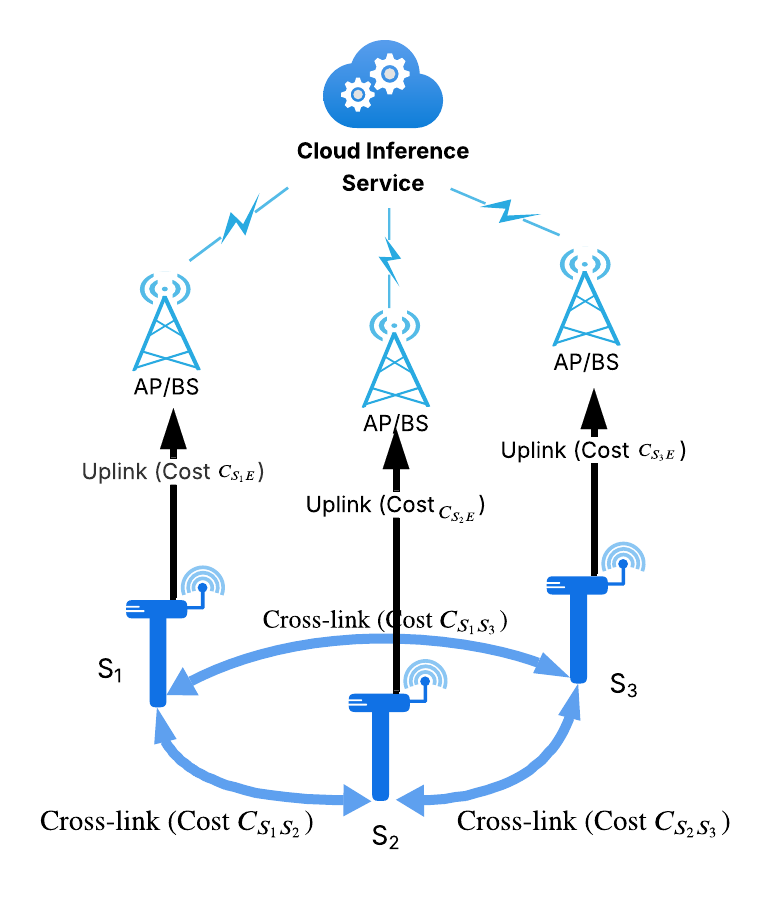}
    \caption{Illustrative diagram of a sample network architecture of a Low-Power Wireless Sensor Network (LPWSN) used in our proposed collective inference model}
    \label{fig:net_diag}
\end{figure}

Examples of this collective, distributed inference can be found in many Wireless Sensor Network use cases. In Fig.~\ref{fig:net_diag}, we illustrate a template of such a use case, which comprises a set of sensing devices, spread over some geographic area that intermittently collect data and perform inference, such as detecting poor air quality or predicting flooding. We assume that the sensing devices can communicate directly with one another via low-cost communication channels, denoted as the cross-link channels in Fig.~\ref{fig:net_diag}.  We also assume access to a centralised cloud inference service via some higher cost uplinks through Access Points or Base Stations. Here, we assume that our task is to perform collective inference over data from all sensing devices. This inference can either be performed locally at the level of the individual sensor(s), and/or by the cloud inference service based on uploaded data from some or all sensors. In practice, this type of network architecture can arise in cases like UAV swarms, where UAVs may be able to utilise short-range communication methods such as WiFi or Bluetooth Low Energy (BLE) alongside longer-range communication channels such as 5G to a Ground Station \cite{nelsonRLBasedEnergyEfficientData2024}.  Other applications where such a scenario may be applicable involve Vehicular Networks, Smart Home networks and wireless sensor networks in Cities that incorporate wireless technologies such as WiFi, LoRa, and NB-IoT \cite{chenCognitiveLPWANIntelligentWireless2019} in a similar network architecture.

Current approaches do not consider the benefits, challenges, and implementation of hybrid frameworks that combine elements of cloud, fog, edge, and distributed inference. Recent research within Multi-Agent Systems has identified the strong potential of such hybrid frameworks \cite{mooreTaxonomyHierarchicalMultiAgent2025} for communication-efficient collective inference. To the best of our knowledge, there has been no detailed exploration or implementation of such a hybrid framework in practice. Here, we consider some simple cases involving sensor networks sensing some Gaussian distributed physical phenomenon, and evaluate the performance of a hybrid inference framework.

Our hybrid framework achieves communication efficiency by operating a dynamic communication strategy based on communication cost optimization. Hence, its performance is strongly dependent on communication cost estimates.  While the experimental evaluations are based on simple estimates of transmission energy costs, our approach can be implemented in practice with more sophisticated cost metrics. In our previous work \cite{nashDistributedClassificationDynamic2025}, we considered a proof-of-concept where we applied such a framework to an Air Quality dataset. In this paper, we take a more general approach, rather than focusing on any individual use case or application. 

Our main contribution in this research is to create the first implementation of such a novel hybrid framework, by integrating peer exchange of sensor data with centralised aggregation and inference to optimize metrics of communication cost. We demonstrate the performance benefits, in terms of accuracy and communication cost, over conventional distributed and centralised approaches.
 This paper is organized as follows.   
 In Section \ref{rel_work}, we define and discuss related concepts to our proposed collective inference framework, and detail the state of the art within each.
 In Section \ref{sys_model}, we define our system model, formalise the problem of collective inference, and define the structure of our proposed framework.
 In Section \ref{results}, we demonstrate realisations of our framework under various scenarios, and compare performance against some relevant baselines.
 Finally, in Section \ref{conclusions}, we summarise our work, and outline the opportunities for future research and development of our proposed collective inference framework. 

\section{Related Work}\label{rel_work}

In the existing literature, there are three broad categories of research that deal with the challenges of collaborative classification, with respect to distributed data sourced from sensing devices (our present focus). These can be classified as:

\begin{enumerate}\label{broad_approach_cats}
    \item \textbf{Cloud, fog, and edge computing (Cloud/Fog Server(s))} methods. These involve the provisioning of strategically placed servers -- either in some remote cloud location (cloud computing), or close to the sensing (edge) devices (fog computing) for efficient computation, with respect to a pre-considered metric.
    \item \textbf{Split computing paradigms (Hierarchical)} that involve performing partial inference on edge, fog, and cloud computing devices. We consider hierarchical split-computing paradigms to be those that involve more than two levels of partial inference. 
    \item \textbf{Fully distributed computing (P2P Comm.)} where data transfer takes place between sensing devices without the use of any dedicated additional computing device, until a \textit{consensus} is reached.
\end{enumerate} 
\ \\
 
Within these broad categories, there are a number of important relevant concepts that we define as follows:

\subsection{Early Exit/Edge}\label{EE}
Early exit methods are typically considered within the hierarchical and other Split Computing frameworks. These enable sensing devices to make direct classification of the data, should they be able to do so with some sufficient pre-determined minimum level of confidence. This can be seen in the work of Shu et al \cite{shuCollaborativeIntrusionDetection2021}, which relies heavily on early exits performed along a hierarchical data flow from the sensing devices through various fog and cloud servers.

\subsection{Agent Based Methods}
Within the contexts of both hierarchical and fully distributed frameworks, there are methods that are \textit{agentic} in nature -- this means that each sensing or processing (cloud/fog server) device makes independent decisions based on the data it receives, without relying on a centralised orchestrator. These decisions are typically taken on a set of actions encompassing options for data transfer, request, or decision reporting. Yuming et al \cite{yumingDecentralizedConsensusInferencebased2025} apply Multi-Agent reinforcement Learning with communication to achieve co-operation between UAV agents based on local observations, received information, and each agent's predictions of their peers' actions. 

\subsection{Dynamic Communication}
There are certain approaches within distributed and Split Computing frameworks that pre-define a data communication schedule between devices and servers that is independent of any observations made at inference time by any of the devices. We define `dynamic communication' frameworks as those that do \textbf{not} feature such a pre-defined schedule, and permit devices to perform actions that are determined solely based on their observed and/or received data at inference time. Yuming et al \cite{yumingDecentralizedConsensusInferencebased2025} again provide a good example of this, where agents make local decisions as to who to communicate with at inference time.

\subsection{Power/Communication Efficient}
Across both centralised and distributed frameworks, different works focus on different metrics of communication or computation cost -- such as latency, bandwidth or battery usage as objective functions for optimizing communication policies or schedules. This choice of metric is driven largely by the application scenario. There are many possible ways to apply this -- for example, in \cite{baccarelliLearningintheFogLiFoDeep2021}, their communication and early exit strategy is driven by minimising the inference delay.

\subsection{Value of Information}\label{VOI}
While less considered in the context of the three broad categories of inference framework mentioned above, there have been numerous cases in the literature that considered methods of maintaining estimates or belief states over data held by peer devices, and methods for estimating the value of this data, called Value of Information. As an example, Zhang et al \cite{zhangCollaborativeFusionDistributed2018}, rely heavily in their work on each sensing agent's estimates of the reliability of each data point and the prediction that it generates.

In Table \ref{tax_tab}, we summarise these works in a taxonomy and comparison of the relevant literature. A key limitation within the literature is that there has been no research that incorporates all of these concepts into a single collective inference framework

\begin{table}[htbp]
\caption{Comparison of Related Works}\label{tax_tab}
\begin{center}
\begin{tabular}{|l|c|c|c|c|c|c|c|c|c|}
\hline
\rot{\textbf{Reference}}\hfill&\rot{Dist. Data} &\rot{Agent Based}&\rot{P2P Comm.}&\rot{Dynamic Comm.}&\rot{Pow./Comm. Efficient}&\rot{Hierarchical}&\rot{Early Exit/Edge}&\rot{Value of Information  }&\rot{Cloud/Fog Server(s)} \\\hline
\cite{qasemMultiAgentSystemCombined2021} &\checkmark &\checkmark & \checkmark &  \checkmark&  \checkmark& &&\checkmark &  \\\hline 
\cite{zhouPeertopeerCollaborativeIntrusion2005}\cite{yumingDecentralizedConsensusInferencebased2025} & \checkmark &\checkmark &\checkmark &\checkmark  &  &  &  &  & \\\hline
\cite{khanCollaborativeSVMClassification2017} & \checkmark &\checkmark &\checkmark &  &\checkmark  &  &  &  & \\\hline
\cite{kazariDecentralizedAnomalyDetection2023} & \checkmark &\checkmark &\checkmark &  &  &  &  &  & \\\hline
\cite{yuWhich2commEfficientCollaborative2025} & \checkmark& & \checkmark &\checkmark  & \checkmark & && &  \\\hline 
\cite{alhakeemDecentralizedBayesianDetection1996} & \checkmark & & &  &  &  &\checkmark  &  & \\\hline
\cite{malkaDecentralizedLowLatencyCollaborative2025} & \checkmark & & \checkmark&  &  &  &\checkmark &   & \\\hline
\cite{zhangCollaborativeFusionDistributed2018} & \checkmark & & & & & &  &\checkmark & \\\hline 
\cite{shuCollaborativeIntrusionDetection2021}& \checkmark & & & & & \checkmark & \checkmark & & \checkmark \\\hline
\cite{baccarelliLearningintheFogLiFoDeep2021} & \checkmark & & & \checkmark & \checkmark & \checkmark & \checkmark & & \checkmark \\\hline 
\cite{zhouPeertopeerCollaborativeIntrusion2005} & \checkmark & & & & \checkmark & && & \checkmark \\\hline 
\cite{nazzalSemiDecentralizedInferenceHeterogeneous2024} &\checkmark & & \checkmark &  &  \checkmark& \checkmark&& &  \checkmark\\\hline 
\cite{shaoTaskOrientedCommunicationMultidevice2023} &\checkmark & &  &  &  \checkmark& \checkmark&& &  \checkmark\\\hline 
\cite{stahlDeeperThingsFullyDistributed2021} & & & \checkmark &  \checkmark&  & && & \checkmark \\\hline 
\cite{liAppealNetEfficientHighlyAccurate2021} & & &  & \checkmark &  & &\checkmark& &  \checkmark\\\hline 
Proposed & \checkmark &\checkmark &\checkmark &\checkmark  &\checkmark  &\checkmark  &\checkmark  &\checkmark  &\checkmark \\\hline
\end{tabular}

\end{center}
\end{table}

\section{System Model and Proposed Framework}\label{sys_model}

\subsection{Notation, Definitions and Assumptions}

Table \ref{tab:notation_and_symbols} defines the basic notation used throughout this paper to define the underlying system model, collective inference task, and our proposed inference framework.

\begin{table}[htbp]
    \centering
    \caption{Notation and Symbols used}
    \begin{tabular}{|c|p{5cm}|}
    \hline
         $Y$ & The global state or outcome being predicted\\\hline
         $N$ & Number of sensing devices, or data sources \\\hline
         $S_i, \;s.t. \;i\in [1,N]$& The random variable describing the data generated at sensor $i$ \\\hline
         $E$& Used to denote a cloud/fog server where joint classification can be performed.\\\hline
         $C_{S_iS_j}, C_{S_iE}$& Cost of communication between sensing devices, or a sensing device and $E$ \\\hline
    \end{tabular}
    \label{tab:notation_and_symbols}
\end{table}

\subsection{Problem Statement}
The problem we are attempting to address is the task of collective inference. We define this as the global prediction/classification of $Y$ at each time step across all sensing devices, either with maximal accuracy within a given \textit{communication cost} budget, or with minimal \textit{communication cost} and a specified minimum overall accuracy.

\subsection{System Model}
 \begin{figure*}[t]
 \subfigure[First component of our proposed framework, corresponding to direct inference on each sensor device.]
    {
        \includegraphics[width=0.27\textwidth]{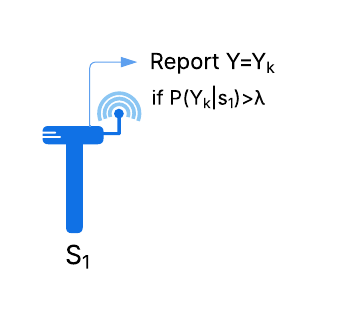}
        \label{fig:direct_exit_arch}
    }\hspace{0.5cm}
      \subfigure[First component of our proposed framework, corresponding to direct inference on each sensor device.]{
        \includegraphics[width=0.33\textwidth]{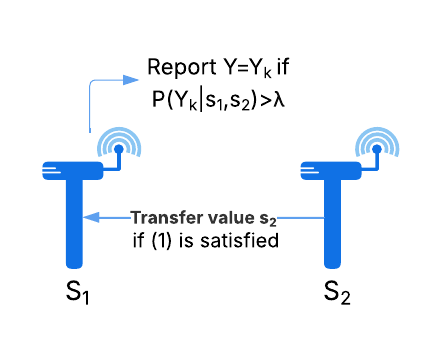}
        \label{fig:req_arch}
    }\hspace{0.5cm}
     \subfigure[Third and final component of our proposed framework, where neither of the previous components are successful.]
    {
        \includegraphics[width=0.27\textwidth]{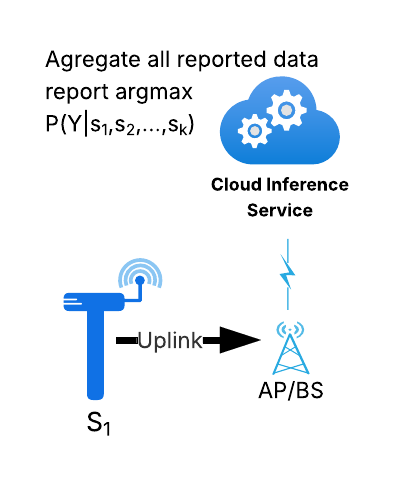}
        \label{fig:offload_arch}
    }
    \caption{Visual representation of the three main aspects of our proposed framework in Section \ref{prop_frame}.}
    \label{fig:visual_policy}
\end{figure*}

Here, we consider a set of $N$ connected sensing devices observing random variables $S_1,S_2,\dots,S_N$. We assume the existence of a centralised cloud/fog server $E$, which has the capacity to collect and process data. We further assume that communication channels exist, each with a specific cost, between some combinations of the sensors (not necessarily a full mesh), and between every sensor and the cloud/fog server. We will denote these costs as $C_{S_iS_j}$ in Joules, where $S_i$ and $S_j$ denote the communicating devices. The specifics of reasoning behind our choice of $C_{S_iS_j}$ are given in Section \ref{cost_def}.

We consider an underlying global hidden state $Y$ across the whole network, which determines the distribution of the sensors' data, i.e. the distribution of $S_i$ varies based on the value of $Y$.

We consider that all sensors process and communicate data synchronously in discrete time steps.

\subsection{Proposed Framework}\label{prop_frame}

We propose addressing the problem stated above by applying a decentralised decision-making strategy independently at each sensor, which dynamically chooses between performing a local prediction of $Y$, data acquisition from a peer device for local inference of $Y$, or data offloading to enable centralised inference.

Formally, we derive and implement policies for the agents (sensors) over the following possible actions, where $\lambda$ is a user-defined confidence threshold that allows for weighting between prediction cost and accuracy. Solutions to these policies then define the decision-making process of our inference framework. The general form of these policies is as follows:

\begin{enumerate}\label{dec_eqns}
    \item Return an early-exit verdict if $P(Y_k|s_i)>\lambda$ for some $k$
    \item Request data from a peer device or devices, if the expected cost (defined in eq. \ref{req_eqn}) of doing so is less than the expected cost of directly offloading data to the cloud/fog server. In the case of requesting up to a single peer device's data, this is determined by: \begin{align}
        C_{S_iE}\,\hat{p}_{ij}+(C_{s_is_j}+C_{S_iE})(1-\hat{p_{ij}})\leq C_{s_iE}, \text{ where } \nonumber\\ \hat{p}_{ij}=\max_{a,j}E_{j}\left[P_j\left(P(Y_k|s_i,S_j)>\lambda\right)\right] \text{ for some } k \label{req_eqn}
    \end{align} If successful, we then need to verify if $P(Y_k|s_i,s_j)>\lambda$ from some $k$ using the requested $s_j$
    \item Otherwise, offload to the cloud/fog server
\end{enumerate}

A summary of the structure of these policies (and therefore, of the resulting framework) is given in Fig.~\ref{fig:visual_policy}. This framework is defined by analytical or approximate solutions to the three terms $P(Y_k|s_i), E_{j}\left[P_j\left(P(Y_k|s_i,S_j)>\lambda\right)\right]$ and $P(Y_k|s_i,s_j)$.

We evaluate this proposed framework by applying it to a selection of exemplar scenarios, as defined by the distributions of $S_i$ and $Y$. This allows us to identify and characterise the situations in which this out- or under-performs conventional collective inference frameworks. To make the analysis more tractable, we limit the evaluation here to the case where $S_i$ are independent Gaussian distributions. This allows us to derive more exact solutions to the decision making criteria defined by (\ref{dec_eqns}), which in turn allows us to reach more robust and confident conclusions about the performance of our framework, albeit in this limited set of scenarios. However, the Gaussian assumption is not a limit of the proposed framework.

Specifically, we consider the following set of distributions:

\begin{enumerate}
    \item $N=2$, Gaussian $S_i$ conditioned on $Y$, all conditionally independent on non-temporal binomial $Y$.
    \item  $N>2$, Gaussian $S_i$ all conditionally independent on non-temporal multinomial $Y$.
\end{enumerate}

Within each scenario, we can derive exact or approximate policies based on solutions to the terms in Section \ref{dec_eqns}. Where possible, we analytically derive distributions of overall accuracy and communication cost over the data parameters and $\lambda$. We also empirically evaluate the emergent behaviour of our framework for different choices of $\lambda$ and data parameters, as well as different network conditions, by applying our policies to repeated random samples from each of the set of distributions.

\subsection{Defining the Cost Metric $C_{AB}$}\label{cost_def}

The aforementioned metric of expected cost of communication $C_{AB}$ forms a critical component of our proposed framework. For this reason, we will implement our framework with metrics of cost that align with conventional collective inference frameworks from the existing literature.

\subsubsection{Conventional Approaches}

Similar issues of optimizing decision-making policies over some communication metrics arise in the areas of split computing and computation resource allocation. Examples of communication metrics commonly used in these contexts are bandwidth utilization, latency, and energy consumption. Authors typically define a primitive set of terms from which to estimate these metrics, which can vary across levels of abstraction. In some cases, the authors will define a set of high-level (and often application-specific) parameters such as wireless transmission power and processing frequency, which are then estimated from device profiling \cite{zengCoEdgeCooperativeDNN2021}. Other methods choose lower-level, more granular primitive terms, which can then be defined directly from principles of the assumed underlying network \cite{kroukaEnergyEfficientModelCompression2021}, \cite{caoJointComputationCommunication2019}. Some authors also extend this principle by considering environmental factors such as path loss \cite{khanQoSEnabledWirelessSplit}.

\subsubsection{Cost Metric Use Case}\label{our_cost_def}

Detailed modelling of specific energy consumption of various types of communication is outside the scope of this research. Hence, we will base our cost metric on the results of the work in \cite{mariniLowPowerWideAreaNetworks2022a}, where Marini, Mikhaylov, et al. compare the performance of Narrowband IoT (NB-IoT) and Long Range Wide Area Network (LoRaWAN). By defining certain energy consumption parameters extracted from the datasheets of the relevant communication modules, and simulating the traffic of various networks using either LoRAWAN or NB-IoT, they were able to determine estimates of the energy consumption and throughput within each simulated network. For simplicity, we use the numerical results from this research by Marini et al. Specifically, we will consider their experimental results corresponding to the simulated energy consumption of end devices separated by 3km, intermittently transmitting 20 byte blocks over a 10 second time window. Their experimental results \cite{mariniLowPowerWideAreaNetworks2022a} compare the energy consumption of an individual device via both LoRaWAN and NB-IoT, over different network sizes (total number of participating end devices).

This leads us to the following hypothetical underlying network scenario for our experimental results and analysis:

\begin{itemize}
    \item Each of our $N$ devices is connected to any other by direct $LoRA$ links, which we assume to have an approximate energy consumption of $1J$ per transmission, to approximately align with data from \cite{mariniLowPowerWideAreaNetworks2022a}.
    \item We also assume that each of the $N$ devices is connected to a cloud inference service, here assuming NB-IoT uplinks to an eNB. We assume that this eNB is subject to sufficient demand from other competing NB-IoT User Equipment, resulting in elevated energy consumption when using the cloud service.
\end{itemize}

Therefore, we will largely hold $C_{S_iS_j}$ fixed across our experimental evaluations, and vary $C_{S_iE}$ from $1J$ to $5J$.

\begin{table}
    \centering
    \caption{Parameters varied across each experiment.}
    \begin{tabular}{|c|p{6cm}|}
    \hline
     Param. &  Explanation\\\hline
        $\delta_\mu$ & $\mu_{s_i|Y=1}-\mu_{s_i|Y=0}$, the degree of separation between distribution means for the same sensor under different states  \\\hline
        $\sigma$ & The standard deviation of all distributions \\\hline
        $\lambda$ & The policy threshold parameter (as per Sec. \ref{prop_frame}) \\\hline
        $C_{S_iE}$ &  The energy expended in communicating a single observation from each sensor to the cloud/fog server (as per Sec. \ref{our_cost_def})\\\hline
        $P_{D}$ & The overall probability that the sensor will be able to directly make a prediction with confidence greater than $\lambda$. This corresponds to the first component of our proposed policy framework.\\\hline
        $P_{RQ}$ & A pair of probabilities ($p_1,p_2$). $p_1$ is the overall probability that the sensor will decide to make a request for $p_2$'s data in the expectation that it has low expected cost -- as defined by the second component of our proposed policy. $p_2$ is the probability of such requests being successful. \\\hline
        $P_E$ & The probability that both the original direct prediction is not sufficiently confident, and a request has an unacceptably high expected cost, resulting in an offload to the cloud/edge. \\\hline
        $E_{C}$ & Defines the sensor's overall expected cost, determined by a weighted Riemann sum over the sensor's density and each point's corresponding expected cost. \\\hline
    \end{tabular}
    \label{tab:main_results_col_exp}
\end{table}

\section{Experiments, Results \& Analysis}\label{results}

We aim to comprehensively evaluate the performance of our framework under each distribution, and this process varies slightly in each case. In our evaluations, we identify the scenarios under which our proposed method out- and under-performs the baselines corresponding to conventional approaches within our data distribution assumptions. 
Both our framework and the baselines are implemented using Python with SciPy and NumPy.

\subsection{Globally Optimal Baseline}
Our proposed framework is based on greedily optimized decisions -- since the overall communication that takes place across the entire network of sensors is determined by decisions based on local observations. Hence, there is a risk that the emergent behaviour of the entire system may contain many redundant communications that can weaken overall performance. In order to determine if, and to what extent, this occurs, we implement an additional baseline based on the solution to an optimal partitioning problem as follows:

\begin{align}\label{globabl_obj_fn}
    \min\sum_{s_1\in S}\left[\sum_{s_2\in S}C_{s_1s_2}\alpha\left(s_1,s_2\right)\right]+C_{s_1E}\beta\left(s_1\right)
\end{align}

\begin{itemize}
    \item $\alpha\left(s_1,s_2\right)\in\{0,1\}$ indicates whether $s_1$ should request data from $s_2$ and perform joint inference on both data. We constrain $\alpha\left(s_i,s_i\right)=0 \;\forall i$, $\alpha\left(s_1,s_2\right)=\alpha\left(s_3,s_2\right) \iff s_1=s_3$, and $\alpha\left(s_1,s_2\right)=\alpha\left(s_1,s_3\right) \iff s_2=s_3$,\\ 
    i.e., each sensor is only involved (either requesting or sending data) in one request -- multiple sensors cannot request a device's data. Further, it is assumed that any sensor that transfers its data does not itself perform local inference after that.
    \item $\beta\left(s_1\right)\in\{0,1\}$ is $1$ if and only if $s_i,s_j\neq s_1 \forall \alpha\left(s_i,s_j\right)$ and $\underset{k}{\arg \max}\; P(Y=y_k|S_1=s_1)<\lambda$, \\
    i.e., this corresponds to offloading data to the edge whenever it is not used for local inference, either jointly with another sensor's data, or by itself.
 \end{itemize}
This defines a globally optimal communication schedule within the same constraints as our proposed framework, while also adding further constraints which should eliminate redundant communications. While impractical in itself, this serves as a useful comparison to determine the extent of redundant communications performed by our proposed framework as the size of the underlying system scales.

\subsubsection{Implementation of Globally Optimal Solution}\label{global_opt_def}

An optimal solution to eq.\ref{globabl_obj_fn} can be found by a recursive backtracking solution. In Table~\ref{tab:empir_results_global_case} (where $C_{S_iS_j}=1J,C_{S_iE}=4J$), we present a set of results describing the performance of the globally optimal baseline. This serves as a reference for comparing the performance of our proposed hybrid framework.

\begin{table}[h]
        \centering
            \caption{Evaluation of the performance of the globally optimized baseline. Explanations of the parameters in each column are available in Table~\ref{tab:main_results_col_exp}. All metrics (other than accuracy) are the averages across all $10\,000$ samples for each set of parameters. The fifth and sixth columns correspond to the average \textit{number} of direct inferences, and requests from a sensor to the other, which leads to a direct inference.}
        \begin{tabular}{|p{0.85cm}|c|c|p{1cm}|p{1cm}|p{1cm}|p{1cm}|}
        \hline
            $\delta_\mu, \sigma$ & $\lambda$  & $C_{S_iE}$ & Acc. (\%) & Avg. \# Direct Decisions  &Avg. \# Successful Requests. &Avg. Cost\\\hline
            (1, 1.5) & 0.75 & 1 & 0.119 & 0.0 & 0.0 & 0.0 \\\hline
            (1, 1.5) & 0.75 & 2 & 0.119 & 0.88 & 0.203 & 2.281 \\\hline
            (1, 1.5) & 0.75 & 4 & 0.119 & 0.88 & 0.203 & 3.682 \\\hline
            (1, 1.5) & 0.85 & 1 & 0.013 & 0.0 & 0.0 & 0.0 \\\hline
            (1, 1.5) & 0.85 & 2 & 0.013 & 0.986 & 0.073 & 2.814 \\\hline
            (1, 1.5) & 0.85 & 4 & 0.013 & 0.986 & 0.073 & 4.643 \\\hline
            (1, 1.5) & 0.95 & 1 & 0.0 & 0.0 & 0.0 & 0.0 \\\hline
            (1, 1.5) & 0.95 & 2 & 0.0 & 0.999 & 0.003 & 2.991 \\\hline
            (1, 1.5) & 0.95 & 4 & 0.0 & 0.999 & 0.003 & 4.983 \\\hline
            (2, 1.5) & 0.75 & 1 & 0.503 & 0.0 & 0.0 & 0.0 \\\hline
            (2, 1.5) & 0.75 & 2 & 0.503 & 0.496 & 0.451 & 1.04 \\\hline
            (2, 1.5) & 0.75 & 4 & 0.503 & 0.496 & 0.451 & 1.585 \\\hline
            (2, 1.5) & 0.85 & 1 & 0.285 & 0.0 & 0.0 & 0.0 \\\hline
            (2, 1.5) & 0.85 & 2 & 0.285 & 0.714 & 0.329 & 1.673 \\\hline
            (2, 1.5) & 0.85 & 4 & 0.285 & 0.714 & 0.329 & 2.631 \\\hline
            (2, 1.5) & 0.95 & 1 & 0.063 & 0.0 & 0.0 & 0.0 \\\hline
            (2, 1.5) & 0.95 & 2 & 0.063 & 0.936 & 0.166 & 2.499 \\\hline
            (2, 1.5) & 0.95 & 4 & 0.063 & 0.936 & 0.166 & 4.062 \\\hline
            (5, 1.5) & 0.75 & 1 & 0.931 & 0.0 & 0.0 & 0.0 \\\hline
            (5, 1.5) & 0.75 & 2 & 0.931 & 0.068 & 0.904 & 0.082 \\\hline
            (5, 1.5) & 0.75 & 4 & 0.931 & 0.068 & 0.904 & 0.095 \\\hline
            (5, 1.5) & 0.85 & 1 & 0.888 & 0.0 & 0.0 & 0.0 \\\hline
            (5, 1.5) & 0.85 & 2 & 0.888 & 0.111 & 0.863 & 0.142 \\\hline
            (5, 1.5) & 0.85 & 4 & 0.888 & 0.111 & 0.863 & 0.173 \\\hline
            (5, 1.5) & 0.95 & 1 & 0.785 & 0.0 & 0.0 & 0.0 \\\hline
            (5, 1.5) & 0.95 & 2 & 0.785 & 0.214 & 0.749 & 0.322 \\\hline
            (5, 1.5) & 0.95 & 4 & 0.785 & 0.214 & 0.749 & 0.43 \\\hline
            (7, 1.5) & 0.75 & 1 & 0.987 & 0.0 & 0.0 & 0.0 \\\hline
            (7, 1.5) & 0.75 & 2 & 0.987 & 0.012 & 0.981 & 0.013 \\\hline
            (7, 1.5) & 0.75 & 4 & 0.987 & 0.012 & 0.981 & 0.013 \\\hline
            (7, 1.5) & 0.85 & 1 & 0.977 & 0.0 & 0.0 & 0.0 \\\hline
            (7, 1.5) & 0.85 & 2 & 0.977 & 0.022 & 0.971 & 0.023 \\\hline
            (7, 1.5) & 0.85 & 4 & 0.977 & 0.022 & 0.971 & 0.024 \\\hline
            (7, 1.5) & 0.95 & 1 & 0.957 & 0.0 & 0.0 & 0.0 \\\hline
            (7, 1.5) & 0.95 & 2 & 0.957 & 0.042 & 0.943 & 0.047 \\\hline
            (7, 1.5) & 0.95 & 4 & 0.957 & 0.042 & 0.943 & 0.052 \\\hline
        \end{tabular}
    \label{tab:empir_results_global_case}
\end{table}

\subsection{$N=2$, Gaussian $S_i$, conditionally independent on binomial $Y$}\label{case_1_exp}

For this trivial 2-sensor scenario, our approach has no practical utility -- nor do other state-of-the-art centralised and distributed frameworks. Transfer of one sensor's data to the other allows for complete joint inference at the lowest possible cost. However, the simplicity of this scenario allows us to determine closed-form analytical solutions to the components of our framework, which lead to useful insights. We use these insights to direct experimentation in more complex and realistic scenarios.

\subsubsection{Theoretical Derivations}\label{theory_results}
Here, we consider Gaussian $S_i$ conditioned on $Y$, all conditionally independent and non-temporal. $Y$ as a non-temporal binomial distribution.

\begin{enumerate}
    \item $P(Y_k|s_i)$ is calculated simply as \begin{equation}\displaystyle
        \frac{f(s_i|y_k)P(Y=y_k)}{\sum_{Y_j} f(s_i|Y_j)}
    \end{equation}

    \item $\hat{p}_{ij}$ is calculated as the probability of the solutions in $s_j$ to \small\begin{align}
       &\left(\frac{1}{\sigma_{j|y_l}^2}-\frac{1}{\sigma_{j|y_k}^2}\right)s_j^2\nonumber+2\left(\frac{\mu_{j|y_k}}{\sigma_{j|y_k}^2}-\frac{\mu_{j|y_l}}{\sigma_{j|y_l}^2}\right)s_j\nonumber\\
       &+\frac{\mu_{j|y_l}}{\sigma_{j|y_l}^2}-\frac{\mu_{j|y_k}}{\sigma_{j|y_k}^2}+2\log\left[\frac{P(y_k|s_i)(1-\lambda)}{P(y_l|s_i)\lambda}\frac{\sigma_{j|l}}{\sigma_{j|k}}\right] > 0 \label{eq:sol_poly_basic_gauss} 
    \end{align}\normalsize 
    \item And $P(Y_k|s_i,s_j)$ is trivially calculated as  \small\begin{equation}\displaystyle
        \underset{k}{\arg \max}\frac{f(s_i|y_k)f(s_j|y_k)P(y_k)}{f(s_i)f(s_j)}
    \end{equation}\normalsize 
\end{enumerate}

\subsubsection{Considering Analytical Metrics of Our Proposed Policy's Performance}\label{vis_basic_g}

For each given set of data parameters of interest, we consider an evenly spaced range of $1000$ values for each $S_1$ and $S_2$, with the upper and lower bounds enclosing close to $100\%$ of the density of each. We then apply of framework (\ref{prop_frame}) across these data. This allows us to map intervals over $s_0$ and $s_1$ to corresponding actions. This allows us to examine the expected behaviour of our proposed framework in various scenarios, and identify where it has a lower cost compared to a pure cloud/edge baseline.

\subsubsection{Visualisations of Analytical Metrics}\label{vis_metrics}
For the following figures, we hold the communication cost parameters constant as  $C_{S_iS_j}=1,C_{S_iE}=4$, and $\lambda=0.75$. This leads to a baseline cost of $2C_{S_iE}=8\text{J}$. In all figures, the blue and orange lines correspond to the densities of each sensor distribution under the two possible states $Y$. The green regions correspond to values where it is possible to make a direct inference for the corresponding x-axis values of $s_i$. The yellow regions correspond to values of $s$ for which the respective sensor performs requests for the other sensor's data. Red regions correspond to values of $s_i$ where neither is possible, and an edge offload is required.

While not rigorous, it is useful for an initial analysis to consider a narrow set of parameters so that we can visually examine the behaviour of our proposed framework. This helps guide our broader and more rigorous evaluations in the following.

\begin{figure}[t]
 \subfigure[Distinct separation of the sensor's distribution between states creates regions trivially handled by independent classifiers.]
    {
        \includegraphics[width=0.22\textwidth]{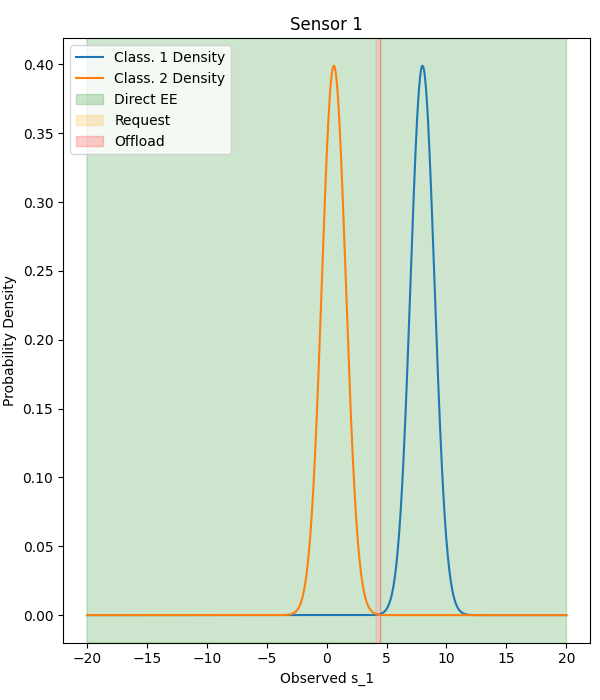}
        \label{fig:bothwell}
    }\hfill
    \subfigure[Case with a small degree of separation between the densities of each sensor for different states. Small (yellow) regions where our policy may be beneficial.]{
        \includegraphics[width=0.22\textwidth]{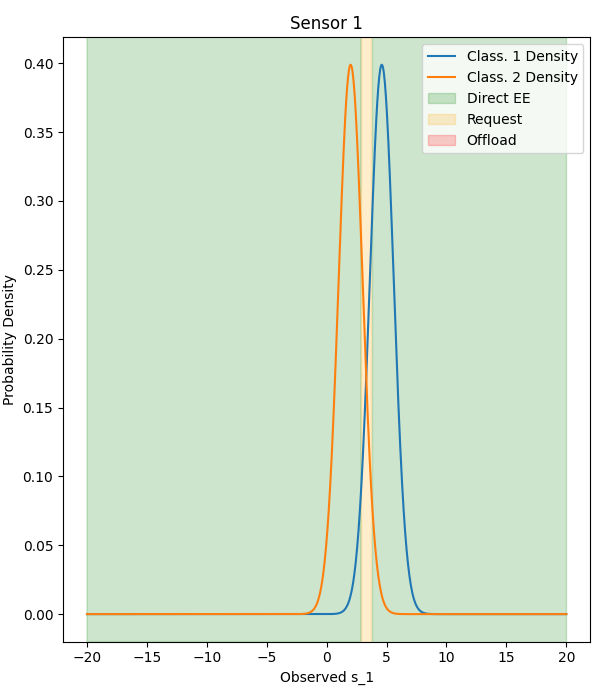}
        \label{fig:modlowvar}
    }\hfill
     \subfigure[Increased variance can create wider regions of overlap and separation, which should increase the utility of our approach.]
    {
        \includegraphics[width=0.22\textwidth]{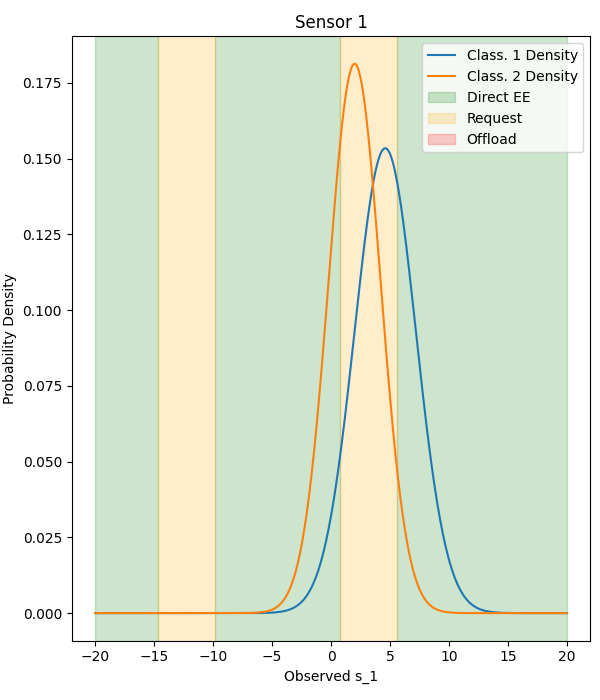}
        \label{fig:modhighvar}
    }\hfill
    \subfigure[Poor separation of the sensor's distribution between states. Creates regions where requests are made, but tend to be unsuccessful.]
    {
        \includegraphics[width=0.22\textwidth]{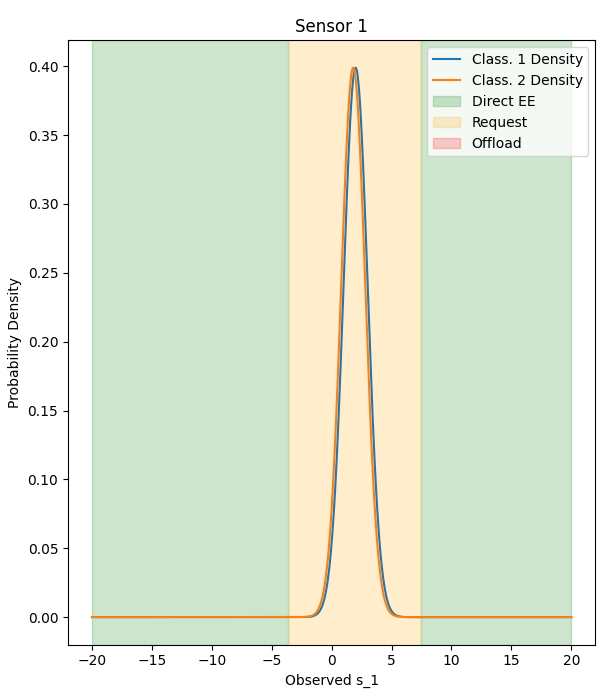}
        \label{fig:bothpoor}
    } 
    \caption{Visualisations of basic scenarios where our framework shows trivial behaviour.}
    \label{fig:basic_visuals}
\label{fig:all_basics}
\end{figure}

As we may intuitively expect, our framework does not perform well on data when the distribution of each sensor's data is either very similar or very distinct between the two possible system states. Fig.~\ref{fig:all_basics} shows the distribution of the sensors' data, under 4 trivial scenarios. For each scenario, we illustrate the distribution of one of the sensors involved. 

\begin{itemize}
    \item Fig.~\ref{fig:bothwell} corresponds to a situation where the sensor distributions are well separated between the two possible $Y$ states. Here, we see that over virtually all values of $S_i$, it is possible to directly infer $Y$ with confidence greater than $\lambda$ in almost all cases (99.89\%). For values of $S_i$ equidistant between the two states, edge offloads are performed, such requests don't appear to lead to confident inference. 
    \item Fig.~\ref{fig:bothpoor} corresponds to a situation with poor separation of sensor values between states. It initially appears that our framework may (counter-intuitively) be able to usefully perform requests. However, while there are nearly 100\% and 99.3\% probabilities of these requests being performed by sensors 1 and 2 respectively, these only have a 1.5\% and 1\% probability of success in practice. This discrepancy is caused by a possibly overly large difference between the costs of cloud/fog and inter-sensor communication. In these scenarios our framework devolves into a conventional cloud computing scenario with some early-exit cost reduction (around 50\% less transmission energy consumed).
    \item Fig.~\ref{fig:modlowvar} and Fig.~\ref{fig:modhighvar} correspond to situations with moderate separation between the state distributions, with low and high variance respectively. Within both, we see regions of overlap between the densities where requests are useful -- and in the case of higher variance, we see wider regions of overlap. For sensor 1 the chance of these requests succeeding is 66\% in the case of Fig.~\ref{fig:modlowvar}, and $30\%$ in the case of Fig.~\ref{fig:modhighvar}. We expect this low success rate is caused by the relatively large disparity between $C_{S_iS_j}$ and $C_{S_iE}$
\end{itemize}

\begin{figure}[htbp]
    \centering
    \includegraphics[width=1\linewidth]{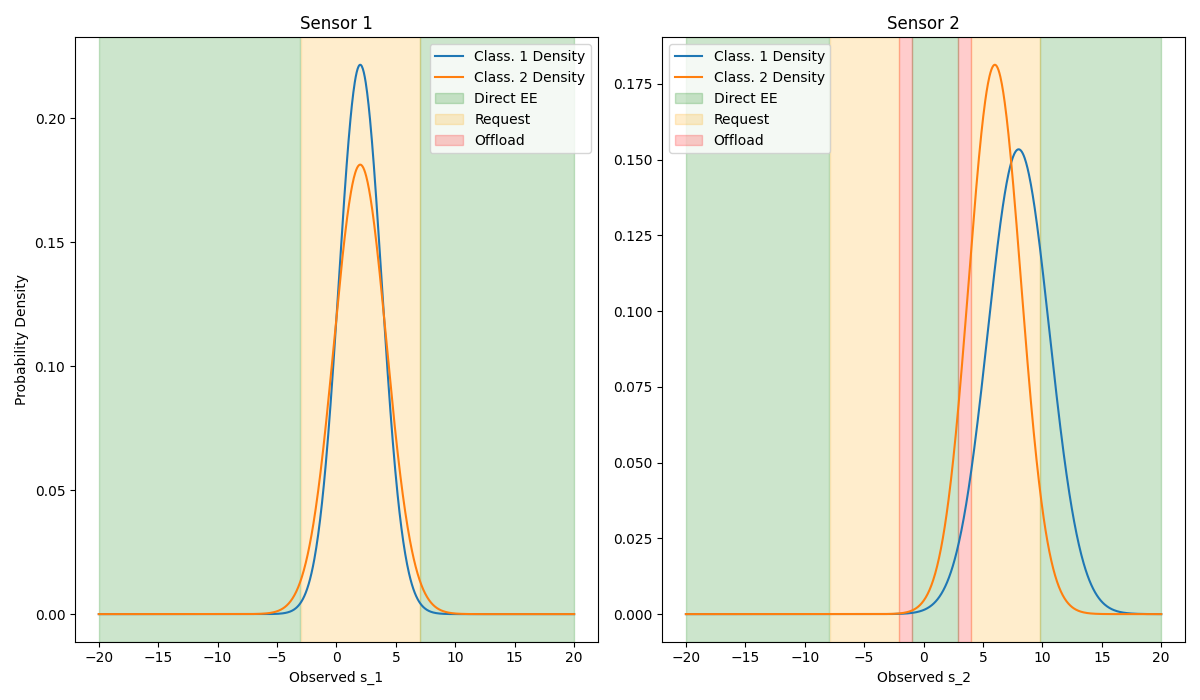}
    \caption{Case where one sensor is poorly separated and one well separated.}
    \label{fig:1well1poor}
\end{figure}

In the particular case where we have one sensor well separated between states, and the other poorly separated, we see a combination of these modes of behaviour to good effect in Fig.~\ref{fig:1well1poor}. Here Sensor 1 is poorly separated, with a $98.6\%$ probability of performing a request with a $12\%$ chance of success. Sensor 2 only performs direct inference in $1.9\%$ of cases, and performs both requests ($73.47\%$ probability) and cloud/fog offloads ($7\%$ probability). These requests only succeed in $7.5\%$ of cases however, again indicating that the discrepancy in costs for inter-sensor and cloud/fog communication is again making a significant contribution. The overall costs (4.9J and 4.1J) cumulatively exceed the cloud baseline by $1J$, as we might expect for such a trivial scenario. We expect cost-effectiveness to be achieved with higher number of sensors, for realistic scenarios.

\subsubsection{Tabulation \& Detailed Analysis of Analytical Results}

In all cases, we assume $C_{S_iS_j}=1J$. For initial experiments and evaluations, we vary the cost by varying $C_{S_iE}$. This leads to the costs (in J) for a sensor-cloud/edge offload being $N\times C_{S_iE}$.

Table \ref{tab:main_results_case_1} gives an overview of how our proposed framework performs over a broad range of parameters of interest.

\paragraph{Separation between states ($\delta_\mu$)}{This defines the degree of separation between the distributions of the sensors' data between states. Where every other experimental parameter is held constant, increasing $\delta_\mu$ leads to decreased expected cost. This makes intuitive sense, as the ability to make confident inference increases with better separation between the distributions. At higher $\delta_\mu$, our framework begins to act as a set of independent classifiers (with direct inference in 98.7\% of cases, and cost approaching 0) as we observed in Fig~\ref{fig:bothwell}. For very low separation ($\delta_\mu\in \{1,2\}$), we see that the expected cost is greater than $C_{S_iE}$, indicating little utility of our framework in these cases. For $\delta_\mu=5$ (and $\delta_\mu=2$, with very high $\lambda$), we can observe a balance between direct inference and often-successful requests. In these cases, our framework is not simply reduced to a set of independent classifiers and has noticeably lower cost than the cloud/fog method.
}

\paragraph{Confidence threshold ($\lambda$)}{As we might intuitively expect, increasing the required confidence threshold has two immediately observable effects -- decreasing the number of direct inferences (leading to increased requests for peer sensors' data), and decreasing the probability that requests lead to successful inference. In the case where there is a moderate degree of separation ($\delta_\mu \in \{2,5,7\}$), increasing $\lambda$ leads to an increasing cost that remains lower than that of the cloud/edge baseline. Further, the instances where we see the highest probability of requests taking place (which is the fundamental novelty of our approach) are instances where $\lambda$ is highest.}

\paragraph{Energy expended in communication from the sensor to the cloud/fog server $C_{S_iE}$}{
As the cost increases, we observe increasing overall costs. However, most critically, we see that these increases in overall cost are much lower than the corresponding increases in $C_{S_iE}$ when all other parameters mentioned above allow for a combination of direct inference and effective requests to take place. This provides a further indication that the utility of our framework increases as the system scales.}

\begin{table}[hbtp]
        \centering
        \caption{Evaluation of the functioning of our proposed framework on the basic case of two Gaussian sensors and a binomial hidden state over a broad range of parameters. Explanations of column headings are given in Table \ref{tab:main_results_col_exp}}
        \begin{tabular}{|c|c|p{0.55cm}|p{0.79cm}|c|p{0.85cm}|p{0.8cm}|}
        \hline
            $\delta_\mu, \sigma$ & $\lambda$  & $d_{S_iE}$ & $P_{D}(S_1)$  &$P_{RQ}(S_1)$&$P_E(S_1)$  & $E_{C}(S_1)$ \\\hline
            (1, 1.5) & 0.75 & 5 & 0.119 & (0.88, 0.203) & 0.0 & 4.782 \\\hline
            (1, 1.5) & 0.75 & 10 & 0.119 & (0.88, 0.203) & 0.0 & 6.597 \\\hline
            (1, 1.5) & 0.75 & 25 & 0.119 & (0.88, 0.203) & 0.0 & 17.839 \\\hline
            (1, 1.5) & 0.85 & 5 & 0.013 & (0.986, 0.073) & 0.0 & 5.896 \\\hline
            (1, 1.5) & 0.85 & 10 & 0.013 & (0.986, 0.073) & 0.0 & 8.265 \\\hline
            (1, 1.5) & 0.85 & 25 & 0.013 & (0.986, 0.073) & 0.0 & 22.935 \\\hline
            (1, 1.5) & 0.95 & 5 & 0.0 & (0.999, 0.003) & 0.0 & 6.264 \\\hline
            (1, 1.5) & 0.95 & 10 & 0.0 & (0.999, 0.003) & 0.0 & 8.845 \\\hline
            (1, 1.5) & 0.95 & 25 & 0.0 & (0.999, 0.003) & 0.0 & 24.827 \\\hline
            (2, 1.5) & 0.75 & 5 & 0.503 & (0.496, 0.451) & 0.0 & 2.185 \\\hline
            (2, 1.5) & 0.75 & 10 & 0.503 & (0.496, 0.451) & 0.0 & 2.891 \\\hline
            (2, 1.5) & 0.75 & 25 & 0.503 & (0.496, 0.451) & 0.0 & 7.26 \\\hline
            (2, 1.5) & 0.85 & 5 & 0.285 & (0.714, 0.329) & 0.0 & 3.51 \\\hline
            (2, 1.5) & 0.85 & 10 & 0.285 & (0.714, 0.329) & 0.0 & 4.751 \\\hline
            (2, 1.5) & 0.85 & 25 & 0.285 & (0.714, 0.329) & 0.0 & 12.439 \\\hline
            (2, 1.5) & 0.95 & 5 & 0.063 & (0.936, 0.166) & 0.0 & 5.238 \\\hline
            (2, 1.5) & 0.95 & 10 & 0.063 & (0.936, 0.166) & 0.0 & 7.262 \\\hline
            (2, 1.5) & 0.95 & 25 & 0.063 & (0.936, 0.166) & 0.0 & 19.799 \\\hline
            (5, 1.5) & 0.75 & 5 & 0.931 & (0.068, 0.904) & 0.0 & 0.173 \\\hline
            (5, 1.5) & 0.75 & 10 & 0.931 & (0.068, 0.904) & 0.0 & 0.19 \\\hline
            (5, 1.5) & 0.75 & 25 & 0.931 & (0.068, 0.904) & 0.0 & 0.295 \\\hline
            (5, 1.5) & 0.85 & 5 & 0.888 & (0.111, 0.863) & 0.0 & 0.3 \\\hline
            (5, 1.5) & 0.85 & 10 & 0.888 & (0.111, 0.863) & 0.0 & 0.34 \\\hline
            (5, 1.5) & 0.85 & 25 & 0.888 & (0.111, 0.863) & 0.0 & 0.585 \\\hline
            (5, 1.5) & 0.95 & 5 & 0.785 & (0.214, 0.749) & 0.0 & 0.679 \\\hline
            (5, 1.5) & 0.95 & 10 & 0.785 & (0.214, 0.749) & 0.0 & 0.819 \\\hline
            (5, 1.5) & 0.95 & 25 & 0.785 & (0.214, 0.749) & 0.0 & 1.684 \\\hline
            (7, 1.5) & 0.75 & 5 & 0.987 & (0.012, 0.981) & 0.0 & 0.027 \\\hline
            (7, 1.5) & 0.75 & 10 & 0.987 & (0.012, 0.981) & 0.0 & 0.028 \\\hline
            (7, 1.5) & 0.75 & 25 & 0.987 & (0.012, 0.981) & 0.0 & 0.032 \\\hline
            (7, 1.5) & 0.85 & 5 & 0.977 & (0.022, 0.971) & 0.0 & 0.048 \\\hline
            (7, 1.5) & 0.85 & 10 & 0.977 & (0.022, 0.971) & 0.0 & 0.05 \\\hline
            (7, 1.5) & 0.85 & 25 & 0.977 & (0.022, 0.971) & 0.0 & 0.06 \\\hline
            (7, 1.5) & 0.95 & 5 & 0.957 & (0.042, 0.943) & 0.0 & 0.1 \\\hline
            (7, 1.5) & 0.95 & 10 & 0.957 & (0.042, 0.943) & 0.0 & 0.106 \\\hline
            (7, 1.5) & 0.95 & 25 & 0.957 & (0.042, 0.943) & 0.0 & 0.145 \\\hline
        \end{tabular}

    \label{tab:main_results_case_1}
\end{table}

\subsubsection{Empirical Measurement of Performance}
Our analytical results allow us to estimate the probability of our proposed framework making local determinations and performing requests, but this does not give us an easy estimate of overall accuracy and cost compared to baseline methods. In order to make such assessments, we apply an implementation of our proposed framework to $10\,000$ random samples from each of the sets of data distributions that we consider.

We define the baseline methods of interest as:

\begin{itemize}
    \item \textbf{Edge/Fog Server Baseline}
All sensors send data to the cloud/fog server (at cost $N\times C_{SE}$), where joint inference is performed.
    \item \textbf{Independent Classifier Baseline} Each sensor makes a local prediction, each of which contributes to form a global decision, for $0$ overall communication cost.
\end{itemize}

In all cases, both for the baselines and our proposed framework, where there are multiple predicted classes, we predict the class with a plurality of predictions, choosing randomly between ties.

\subsubsection{Empirical Results \& Analysis of Emergent Behaviour}\label{case_1_emergant_behaviour}

In Table \ref{tab:empir_results_case_1}, we consider measures of cost and accuracy for the same sets of parameters we consider for our analytical results in Table \ref{tab:main_results_case_1}. While there is some variation, we see that the results and observations of the empirical experimentation align with those of our analytical metrics in Table \ref{tab:main_results_case_1}.

Due to the small scale ($N=2$) of this scenario, we see that the accuracy of the independent classifier baseline is very high. However, in general we see that our proposed framework achieves an accuracy that is closer to that of the cloud/fog baseline, with a corresponding (often vastly) lower average cost. This effect is seemingly caused by performing direct inferences -- this is clearly observable in the cases where the average number of direct decisions is $\ge1$.

Of more interest to us are cases with a noticeable average number of successful requests ($>0.1$) and a corresponding low average number of direct decisions ($<1$). In these cases, results are mixed -- we see numerous cases where the average cost is less than the cloud/fog baseline. It appears that we see most utility for lower $\delta_\mu$ and high $\lambda$, and accuracy remains consistently higher than the independent classifier baseline. This suggests that poor separation of sensor distributions between states may be the strongest deciding factor that determines the applicability of our framework. Increasing $C_{S_iE}$ increases the utility of our framework. However, under distributions where our method shows utility, we see a benefit over all choices of this parameter. We expect that, so long as the cost $C_{SE}>C_{SS}$, there will be some utility to our framework.

\begin{table}[h]
        \centering
            \caption{Empirical results obtained by simulating the proposed framework on randomly sampled data in the simplest case -- $N=2, Y\in\{0,1\},\sigma=1.5$. All metrics (other than accuracy) are the averages across all $10\,000$ samples for each set of parameters. The fifth and sixth columns correspond to the average \textit{number} of direct inferences, and requests from a sensor to the other, which leads to a direct inference.}
        \begin{tabular}{|p{1.1cm}|p{0.65cm}|p{0.7cm}|p{0.8cm}|p{0.55cm}|p{0.6cm}|p{0.55cm}|p{0.6cm}|}
        \hline
            $\delta_\mu,\lambda$, $C_{S_iE}$ & Acc. (\%) & Avg. \# Direct Decisions  &Avg. \# Successful Requests. &Avg. Cost & Cloud Acc. & Cloud Avg. Cost & Ind. Acc \\\hline
            2, 0.75, 2 & 0.804 & 1.01 & 0.56 & 1.85 & 0.826 & 4.0 & 0.747 \\\hline
            2, 0.75, 4 & 0.805 & 1.01 & 0.55 & 2.73 & 0.827 & 8.0 & 0.748 \\\hline
            2, 0.75, 6 & 0.807 & 1.01 & 0.56 & 3.57 & 0.827 & 12.0 & 0.749 \\\hline
            2, 0.85, 2 & 0.808 & 0.57 & 0.6 & 3.07 & 0.828 & 4.0 & 0.749 \\\hline
            2, 0.85, 4 & 0.808 & 0.58 & 0.6 & 4.72 & 0.828 & 8.0 & 0.748 \\\hline
            2, 0.85, 6 & 0.806 & 0.57 & 0.6 & 6.39 & 0.829 & 12.0 & 0.748 \\\hline
            2, 0.95, 2 & 0.819 & 0.13 & 0.44 & 4.74 & 0.826 & 4.0 & 0.747 \\\hline
            2, 0.95, 4 & 0.819 & 0.13 & 0.44 & 7.59 & 0.827 & 8.0 & 0.747 \\\hline
            2, 0.95, 6 & 0.819 & 0.12 & 0.44 & 10.51 & 0.827 & 12.0 & 0.747 \\\hline
            5, 0.75, 2 & 0.952 & 1.86 & 0.0 & 0.27 & 0.991 & 4.0 & 0.952 \\\hline
            5, 0.75, 4 & 0.952 & 1.87 & 0.0 & 0.54 & 0.991 & 8.0 & 0.952 \\\hline
            5, 0.75, 6 & 0.953 & 1.86 & 0.0 & 0.81 & 0.991 & 12.0 & 0.951 \\\hline
            5, 0.85, 2 & 0.954 & 1.78 & 0.0 & 0.45 & 0.99 & 4.0 & 0.952 \\\hline
            5, 0.85, 4 & 0.954 & 1.78 & 0.02 & 0.82 & 0.991 & 8.0 & 0.953 \\\hline
            5, 0.85, 6 & 0.967 & 1.77 & 0.14 & 0.67 & 0.991 & 12.0 & 0.952 \\\hline
            5, 0.95, 2 & 0.956 & 1.58 & 0.06 & 0.79 & 0.99 & 4.0 & 0.952 \\\hline
            5, 0.95, 4 & 0.976 & 1.58 & 0.33 & 0.75 & 0.99 & 8.0 & 0.952 \\\hline
            5, 0.95, 6 & 0.983 & 1.58 & 0.36 & 0.8 & 0.991 & 12.0 & 0.951 \\\hline
            7, 0.75, 2 & 0.99 & 1.97 & 0.0 & 0.05 & 0.999 & 4.0 & 0.99 \\\hline
            7, 0.75, 4 & 0.99 & 1.97 & 0.0 & 0.1 & 1.0 & 8.0 & 0.99 \\\hline
            7, 0.75, 6 & 0.99 & 1.97 & 0.0 & 0.15 & 1.0 & 12.0 & 0.99 \\\hline
            7, 0.85, 2 & 0.99 & 1.96 & 0.0 & 0.09 & 1.0 & 4.0 & 0.99 \\\hline
            7, 0.85, 4 & 0.99 & 1.96 & 0.0 & 0.17 & 0.999 & 8.0 & 0.99 \\\hline
            7, 0.85, 6 & 0.99 & 1.96 & 0.0 & 0.25 & 1.0 & 12.0 & 0.99 \\\hline
            7, 0.95, 2 & 0.99 & 1.91 & 0.0 & 0.17 & 0.999 & 4.0 & 0.99 \\\hline
            7, 0.95, 4 & 0.99 & 1.91 & 0.0 & 0.35 & 1.0 & 8.0 & 0.99 \\\hline
            7, 0.95, 6 & 0.99 & 1.92 & 0.0 & 0.51 & 1.0 & 12.0 & 0.99 \\\hline
        \end{tabular}

    \label{tab:empir_results_case_1}
\end{table}

\subsubsection{Final Analysis}
While this case of two Gaussian sensors with a binomial hidden state is trivial, it provides good motivation for further experiments. We see clear potential for our framework, particularly where there is poor separation between the means of the distributions of the sensors' data between states. This provides useful context and a robust foundation for our analysis of the Gaussian multi-sensor ($>2$) case, for which we will only be able to derive approximate solutions to our framework.
\subsection{$N > 2$, Gaussian $S_i$, conditionally independent on multinomial $Y$}\label{sec:full_multi_results}

This case, while still simplistic, describes a more interesting and potentially useful scenario.

\subsubsection{Implementation of Our Policy}

Unlike the case of Section~\ref{case_1_exp}, it is not possible to derive complete analytical solutions to every component of our proposed framework defined in \ref{prop_frame} under this data context. Specifically, the challenge arises in the computation of $P_j\left(P(Y_k|s_i,S_j)>\lambda\right)$ from Equation \ref{req_eqn}.

To obtain an approximate solution, we consider:

\begin{align}
    P_{S_j}\left[f(s_j|Y=y_k)\frac{P(Y=y_k|s_i)}{\lambda}\right.\nonumber\\
    \left.-\sum_{l\in dom(Y)}f(s_j|Y=y_l)P(Y=y_l|s_i)>0\right]\label{eq:best_approx_for_multi_gauss}
\end{align}

We obtain an approximation by taking a large uniform sample over a wide range $S_j$ that allows us to directly estimate the range of values, and thus probability in $S_j$, of satisfying the inequality in Equation~\ref{eq:best_approx_for_multi_gauss}. In practice, a similar result can be obtained by applying a combination of multimodal optimization and root-finding strategies. The remaining components of \ref{dec_eqns} are implemented directly.

\subsubsection{Validation Against Previous Results}
In order to confirm the accuracy of this approximation, we evaluate this approach on the trivial Gaussian, 2-sensor, binomial $Y$ approach from Section \ref{case_1_exp}. The equivalent results from Table~\ref{tab:empir_results_case_1} all differ by less than $3.5\%$ in terms of accuracy. The average cost is lower or equal under all instances under our heuristic solution. This provides a positive indication that this approximation serves as an appropriate and effective solution.

\subsubsection{Evaluating a 4-Sensor, 2-Class Scenario}
Here, we consider a more interesting specific scenario (under condition $C_{S_iS_j}=1J$ as before). The results of our evaluation of this scenario are shown in Table~\ref{tab:empir_results_case_2a}, over the same metrics as we saw in Table~\ref{tab:empir_results_case_1}. In this case (and for further cases where we increase the number of predicted classes), there are two significant changes that need to be made to the implementation of our framework:

\begin{enumerate}
    \item Aggregation of multiple predictions (either made by the Cloud/Edge Server, or sensors) is performed by randomly selecting from the classes that are predicted with a joint plurality (or simply the class with a plurality of predictions, if one exists).
    \item The means of the Gaussian distributions of each sensor's data under different classes are separated by increments of $\delta_\mu$, i.e., the distributions for sensor $S_i$ are described by: \begin{align}
        &S_i|Y=0 \sim N(0, \sigma),S_i|Y=1 \sim N(\delta_\mu, \sigma),\nonumber\\&S_i|Y=2 \sim N(2\delta_\mu, \sigma),\dots
    \end{align} 
\end{enumerate}

Our observations here are broadly consistent in character with those in Section~\ref{case_1_emergant_behaviour}, however, the underlying scenario is more realistic and suggest real-world utility for our approach. Here, our proposed framework achieves an accuracy that is always between those of the independent and Cloud/Fog baselines. We can observe that the disparity between the accuracies of Cloud/Edge baseline and Independent baseline is greatest when $\delta_\mu$ is low. In all cases, our proposed framework's accuracy is closer to that of the cloud/fog baseline than the Independent baseline -- while attaining lower cost.

For very high values of $\delta_\mu$, we see costs close to 0 -- this corresponds to the uninteresting case where our framework devolves into a series of independent classifiers. In more interesting cases, with lower $\delta_\mu$, other factors determine cost. With lower values of $C_{\S_iE}$, we see lower overall costs. However, all other parameters being held equal, it seems that the performance of our framework is quite stable in terms of $C_{S_iE}$. For a given set of parameters, varying $C_{S_iE}$ leads to little change in the average number of direct inferences and successful requests. Changes in $\lambda$, on the other hand, have a more pronounced effect. Where $\lambda$ increases with all other parameters held constant, we see a reduction in the number of direct inferences. We can also observe that in certain instances -- such as when $\delta_\mu\in\{2,5\}$ -- this ($\lambda=0.85$) can lead to an increase in the average number of successful requests.

Overall, these suggest that, largely driven by the degree of separation between the distributions of the sensors' data between different states, and controlled by choice of parameter $\lambda$, our framework is able to achieve a potentially useful compromise between the low communication cost of a distributed baseline, and the high accuracy of a centralised baseline.

\begin{table}[h]
        \centering
            \caption{Empirical results obtained by simulating our proposed framework on randomly sampled data in the case of $N=4, Y\in\{0,1,2,3\}, \sigma=1.5$. All metrics (other than accuracy) are the averages across all $10\,000$ samples for each set of parameters. The fifth and sixth columns correspond to the average \textit{number} of direct inferences, and requests from a sensor to the other, which leads to a direct inference.}
        \begin{tabular}{|p{1.1cm}|p{0.65cm}|p{0.7cm}|p{0.8cm}|p{0.55cm}|p{0.6cm}|p{0.55cm}|p{0.6cm}|}
        \hline
            $\delta_\mu$, $\lambda$, $C_{S_iE}$ & Acc. (\%) & Avg. \# Direct Decisions  &Avg. \# Successful Requests. &Avg. Cost & Cloud Acc. & Cloud Avg. Cost & Ind. Acc \\\hline
            2, 0.75, 2 & 0.786 & 1.02 & 0.42 & 1.84 & 0.829 & 4.0 & 0.748 \\\hline
            2, 0.75, 4 & 0.805 & 1.01 & 0.56 & 2.73 & 0.827 & 8.0 & 0.748 \\\hline
            2, 0.75, 6 & 0.806 & 1.01 & 0.56 & 3.56 & 0.828 & 12.0 & 0.747 \\\hline
            2, 0.85, 2 & 0.78 & 0.58 & 0.32 & 2.8 & 0.828 & 4.0 & 0.747 \\\hline
            2, 0.85, 4 & 0.807 & 0.58 & 0.6 & 4.71 & 0.829 & 8.0 & 0.746 \\\hline
            2, 0.85, 6 & 0.806 & 0.58 & 0.6 & 6.37 & 0.826 & 12.0 & 0.746 \\\hline
            2, 0.95, 2 & 0.815 & 0.13 & 0.15 & 3.7 & 0.827 & 4.0 & 0.747 \\\hline
            2, 0.95, 4 & 0.812 & 0.13 & 0.34 & 7.01 & 0.828 & 8.0 & 0.75 \\\hline
            2, 0.95, 6 & 0.815 & 0.13 & 0.39 & 10.08 & 0.827 & 12.0 & 0.748 \\\hline
            5, 0.75, 2 & 0.973 & 1.87 & 0.13 & 0.15 & 0.991 & 4.0 & 0.952 \\\hline
            5, 0.75, 4 & 0.974 & 1.87 & 0.13 & 0.17 & 0.991 & 8.0 & 0.953 \\\hline
            5, 0.75, 6 & 0.973 & 1.86 & 0.13 & 0.19 & 0.991 & 12.0 & 0.952 \\\hline
            5, 0.85, 2 & 0.98 & 1.78 & 0.2 & 0.27 & 0.991 & 4.0 & 0.952 \\\hline
            5, 0.85, 4 & 0.981 & 1.78 & 0.2 & 0.31 & 0.99 & 8.0 & 0.953 \\\hline
            5, 0.85, 6 & 0.98 & 1.78 & 0.2 & 0.36 & 0.99 & 12.0 & 0.952 \\\hline
            5, 0.95, 2 & 0.984 & 1.58 & 0.36 & 0.54 & 0.991 & 4.0 & 0.952 \\\hline
            5, 0.95, 4 & 0.984 & 1.58 & 0.36 & 0.66 & 0.991 & 8.0 & 0.953 \\\hline
            5, 0.95, 6 & 0.984 & 1.58 & 0.36 & 0.78 & 0.991 & 12.0 & 0.952 \\\hline
            7, 0.75, 2 & 0.995 & 1.97 & 0.03 & 0.03 & 1.0 & 4.0 & 0.99 \\\hline
            7, 0.75, 4 & 0.995 & 1.97 & 0.03 & 0.03 & 1.0 & 8.0 & 0.991 \\\hline
            7, 0.75, 6 & 0.995 & 1.97 & 0.02 & 0.03 & 1.0 & 12.0 & 0.99 \\\hline
            7, 0.85, 2 & 0.996 & 1.96 & 0.04 & 0.04 & 0.999 & 4.0 & 0.989 \\\hline
            7, 0.85, 4 & 0.996 & 1.96 & 0.04 & 0.05 & 0.999 & 8.0 & 0.99 \\\hline
            7, 0.85, 6 & 0.996 & 1.96 & 0.04 & 0.05 & 1.0 & 12.0 & 0.99 \\\hline
            7, 0.95, 2 & 0.998 & 1.91 & 0.08 & 0.09 & 1.0 & 4.0 & 0.99 \\\hline
            7, 0.95, 4 & 0.998 & 1.92 & 0.08 & 0.09 & 0.999 & 8.0 & 0.99 \\\hline
            7, 0.95, 6 & 0.998 & 1.91 & 0.08 & 0.1 & 1.0 & 12.0 & 0.99 \\\hline
        \end{tabular}

    \label{tab:empir_results_case_2a}
\end{table}

\subsection{$N > 2$, Gaussian $S_i$, conditionally independent on multinomial $Y$ -- using a more efficient approximate solution}\label{sec:huer_approach}

In Section \ref{sec:full_multi_results}, our method of solving (\ref{eq:best_approx_for_multi_gauss}) using a uniform sample over the sensor's domain is not practical for any realistic scenario. For this reason, we propose a more computationally efficient approximation to
(\ref{eq:best_approx_for_multi_gauss}).
\subsubsection{Our Proposed Heuristic}\label{heur_algo}
We can observe in (\ref{eq:best_approx_for_multi_gauss}), that we can expect solutions (if they exist) in $s_j$ to likely be around $s_j=\mu_{jk}$. This leads us to the following heuristic solution:

\small\begin{align}
    \text{Let } f(s_j) = f(s_j|Y=y_k)\frac{P(Y=y_k|s_i)}{\lambda}\nonumber\\
    -\sum_{l\in dom(Y)}f(s_j|Y=y_l)P(Y=y_l|s_i),\\
    x = \mu_{s_jk}
\end{align}\normalsize 

\begin{enumerate}
    \item If $x>0$, perform ascent and descent in fixed discrete steps (passing over any local maxima) to find the roots of $f$ (or an asymptote at $f(s_i)=0$ is detected)
    \item If $x<0$, perform ascent until a root is reached. Continue performing ascent and descent in fixed discrete steps past any local maxima, until a root or asymptote is found
\end{enumerate}

When a root is detected, the value is estimated by weighted interpolation between the upper and lower bounds.

The approximate solution to (\ref{eq:best_approx_for_multi_gauss}) is then the probability in $S_j$ of the interval described by these roots.

\begin{table}[h]
        \centering
            \caption{Performance of the heuristic algorithm from Section-\ref{heur_algo} under the same conditions as the uniform sampling method in Table~\ref{tab:empir_results_case_2a}.}
        \begin{tabular}{|p{1.2cm}|p{1cm}|p{1.2cm}|p{1.3cm}|p{1cm}|}
        \hline
            $\delta_\mu$, $\lambda$, $C_{S_iE}$ & Acc. (\%) & Avg. \# Direct Decisions  &Avg. \# Successful Requests. &Avg. Cost\\\hline
            2, 0.75, 2 & 0.76 & 1.01 & 0.1 & 1.93 \\\hline
            2, 0.75, 4 & 0.793 & 1.01 & 0.49 & 2.84 \\\hline
            2, 0.75, 6 & 0.806 & 1.01 & 0.55 & 3.61 \\\hline
            2, 0.85, 2 & 0.777 & 0.58 & 0.17 & 2.78 \\\hline
            2, 0.85, 4 & 0.787 & 0.58 & 0.44 & 4.84 \\\hline
            2, 0.85, 6 & 0.8 & 0.57 & 0.54 & 6.56 \\\hline
            2, 0.95, 2 & 0.813 & 0.13 & 0.13 & 3.7 \\\hline
            2, 0.95, 4 & 0.814 & 0.13 & 0.27 & 7.01 \\\hline
            2, 0.95, 6 & 0.814 & 0.13 & 0.32 & 10.14 \\\hline
            5, 0.75, 2 & 0.969 & 1.86 & 0.11 & 0.17 \\\hline
            5, 0.75, 4 & 0.975 & 1.87 & 0.13 & 0.16 \\\hline
            5, 0.75, 6 & 0.974 & 1.86 & 0.13 & 0.19 \\\hline
            5, 0.85, 2 & 0.973 & 1.78 & 0.18 & 0.29 \\\hline
            5, 0.85, 4 & 0.98 & 1.78 & 0.2 & 0.31 \\\hline
            5, 0.85, 6 & 0.98 & 1.78 & 0.2 & 0.35 \\\hline
            5, 0.95, 2 & 0.974 & 1.58 & 0.32 & 0.58 \\\hline
            5, 0.95, 4 & 0.983 & 1.58 & 0.36 & 0.67 \\\hline
            5, 0.95, 6 & 0.984 & 1.58 & 0.36 & 0.79 \\\hline
            7, 0.75, 2 & 0.994 & 1.97 & 0.02 & 0.03 \\\hline
            7, 0.75, 4 & 0.995 & 1.97 & 0.03 & 0.03 \\\hline
            7, 0.75, 6 & 0.995 & 1.97 & 0.03 & 0.03 \\\hline
            7, 0.85, 2 & 0.996 & 1.96 & 0.04 & 0.05 \\\hline
            7, 0.85, 4 & 0.996 & 1.96 & 0.04 & 0.05 \\\hline
            7, 0.85, 6 & 0.996 & 1.96 & 0.04 & 0.05 \\\hline
            7, 0.95, 2 & 0.997 & 1.91 & 0.08 & 0.09 \\\hline
            7, 0.95, 4 & 0.998 & 1.92 & 0.08 & 0.1 \\\hline
            7, 0.95, 6 & 0.998 & 1.92 & 0.08 & 0.1 \\\hline
        \end{tabular}

    \label{tab:empir_results_case_2b}
\end{table}

\subsubsection{Validation of our heuristic against previous results}
In Table~\ref{tab:empir_results_case_2b}, we compare the results of performing the same experiments as under Table~\ref{tab:empir_results_case_2a}, which is based on the uniform-sampling approach. The average number of successful requests (the most important metric to compare here) appears more varied -- in the vast majority of cases, the variation is under a small amount. Accuracy appears consistent, varying less than $3.5\%$ from Table~\ref{tab:empir_results_case_2a}. Cost is much more variable (up to 25\% in cases where the absolute value is small $<0.1J$), but generally appears consistent.

\subsection{Scalability}

Based on our experimental insights into the general characteristics of our proposed framework, we consider its scalability over a much broader range of data. We base these experiments on our heuristic solution defined in \ref{sec:huer_approach} since this is the implementation that is most relevant for real-world applications.

\subsection{Scaling Number of Sensors}

Fig.~\ref{fig:scaling_with_classes} shows the results of scaling $N$, where $C_{S_iE}=4J,C_{S_iS_j}=1J,\lambda=0.85,\sigma=1.5,\delta_\mu=1$. As the number of sensors scales, we see very stable linear growth in the average number of direct inferences, and the number of requests that lead to inferences.  As expected, with an initial cost lower than the centralised baseline, growth in cost is linear in $N$ at a lower rate than the baseline. In this case, cost remains only moderately lower than the cloud baseline. As we observe from Table \ref{tab:empir_results_case_2b}, and in Fig.~9, $\lambda$ and $\delta_\mu$ contribute noticeably to the degree of difference between the cost difference between the centralised baseline and our proposed framework. Improvements in accuracy begin to show diminishing returns at very high $N$, as we could expect with such a large classifier over so many features. As expected, the accuracy of our proposed framework forms a trade-off between that of the independent and centralised baselines.

\begin{figure}
    \centering
    \includegraphics[width=\linewidth]{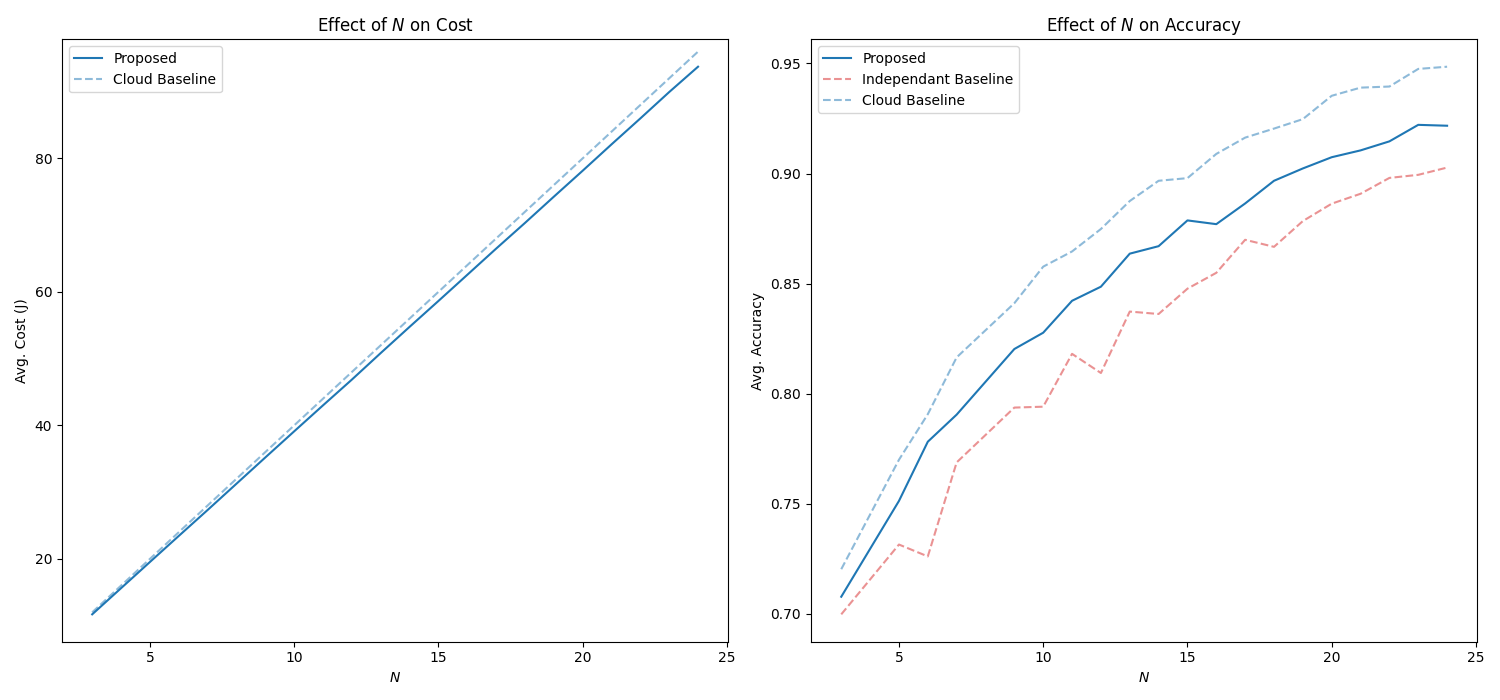}
    \caption{Visualising the effect of increasing $N$ on overall cost and accuracy. }
    \label{fig:scaling_with_n}
\end{figure}

\subsection{Scaling of Parameters}
\paragraph{Scaling $\delta_{\mu}$}

\begin{figure}
    \centering
    \includegraphics[width=0.9\linewidth]{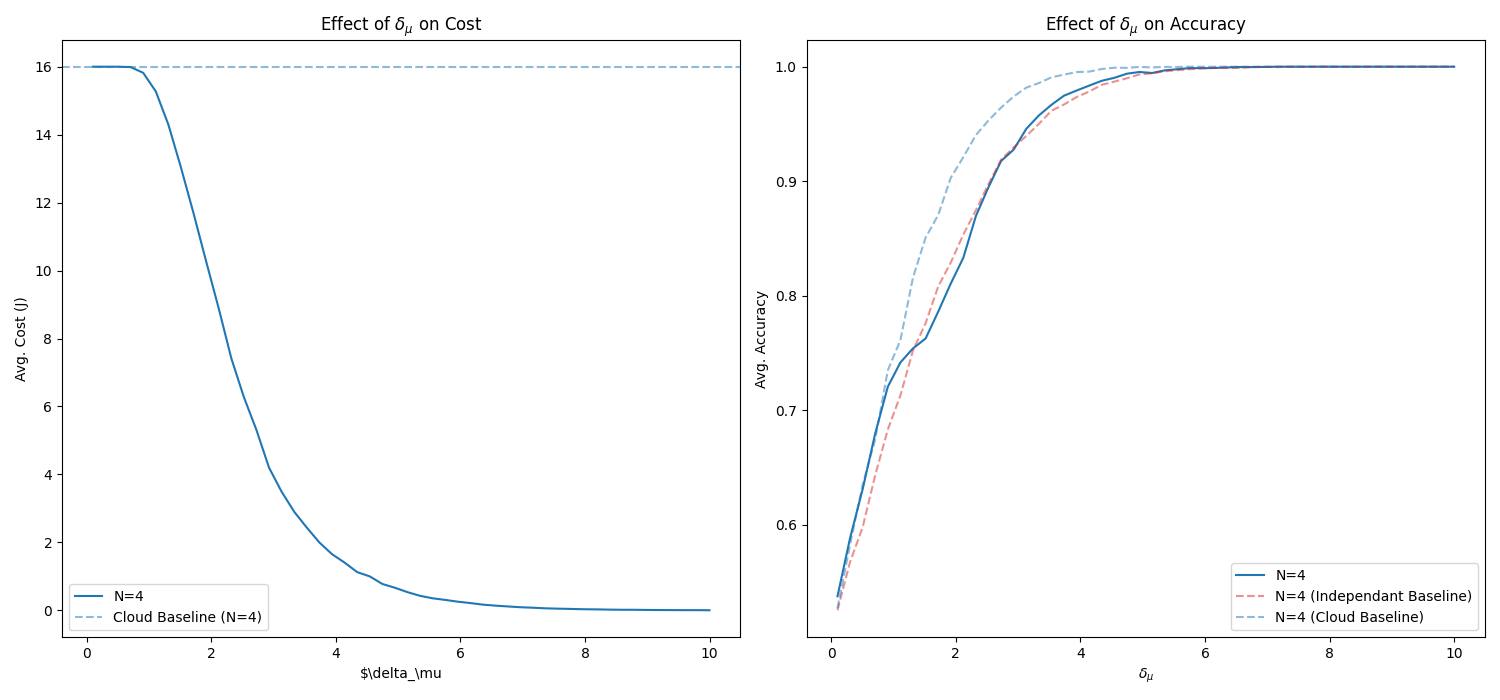}
    \includegraphics[width=0.9\linewidth]{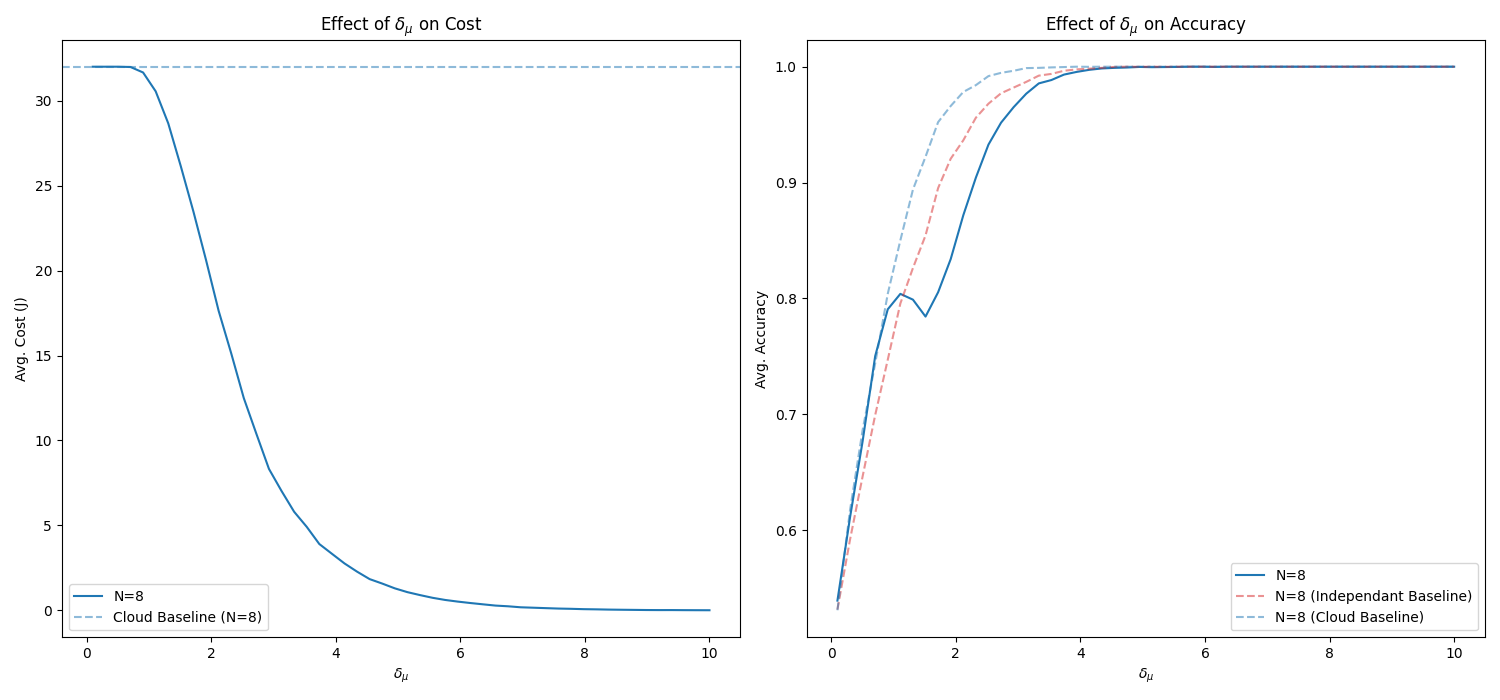}
    \caption{Visualising the effect of increasing $\delta_\mu$ on overall cost and accuracy. N=4 top, N=8 bottom.}
    \label{fig:scaling_with_dmu}
\end{figure}

Fig.~\ref{fig:scaling_with_dmu} demonstrates the effect of varying $\delta_\mu$ with all other parameters fixed as $C_{S_iE}=4J,C_{S_iS_j}=1J,\lambda=0.85,\sigma=1.5$. This provides a direct illustration of the behaviour of our proposed framework as a trade-off between distributed and centralised inference approaches. Reinforcing our observations from Section \ref{vis_metrics}, we see that for very low $\delta_\mu$, where separation between sensor distributions between states is poor, behaviour and thus cost devolves to that of a cloud-based baseline. Similarly, for very high separation, behaviour and cost devolves to a set of independent classifiers. Accuracy increases accordingly, with higher separation leading to a simpler classification task.

Overall, this demonstrates the existence of a critical region in $\delta_\mu$ (given all other parameters), where our framework has utility -- in the case of Fig.~\ref{fig:scaling_with_dmu},  for $N=$ this can be seen where $\delta_\mu$ is between 1 and 1.5. In these instances, our framework will not devolve to either of the baseline approaches and will be able to achieve a level of accuracy approaching the centralised baseline, at lower cost, while also having noticeably higher accuracy than the independent baseline. We expect that the degree of separation between the accuracies of the cloud and independent baselines will have a large impact on the performance of our approach in the critical region. 

Interestingly, in Fig.~\ref{fig:scaling_with_dmu}, where $N=8$, we see that outside this critical region, the overall accuracy of our framework is lower than the independent baseline. This is possibly due to a loss of regularisation effect created by independent classifiers.

\paragraph{Scaling Cost Parameters}

\begin{figure}
    \centering
    \includegraphics[width=\linewidth]{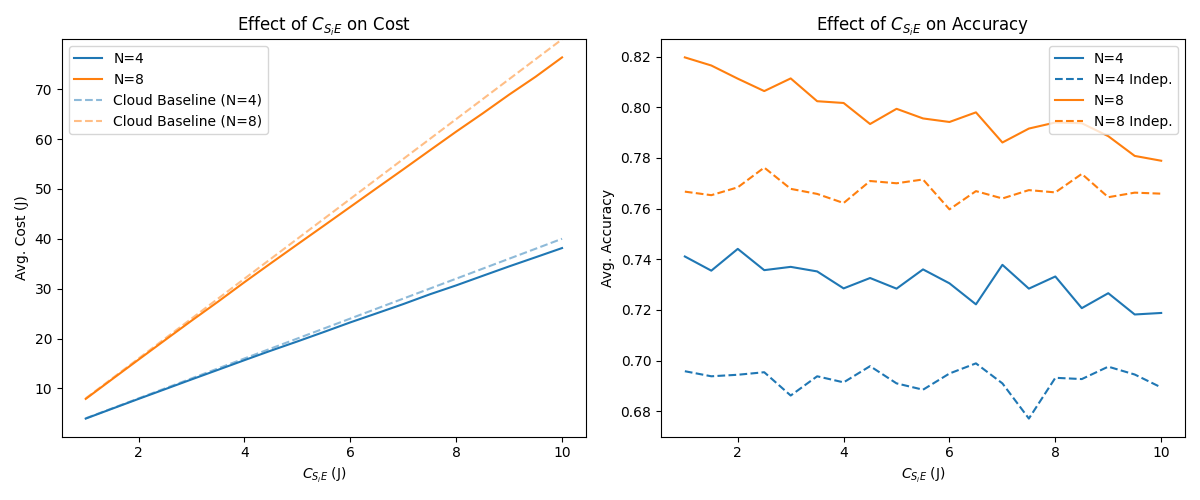}
    \caption{Visualisation of the effect of increasing $C_{S_iE}$ on overall cost and accuracy (assuming $C_{S_iS_j}=1J$. We see cost savings over the cloud baseline increase as the cloud becomes increasingly expensive for offloads. For very low costs, we see an interesting dip in accuracy.}
    \label{fig:scaling_with_cse}
\end{figure}

Fig.~\ref{fig:scaling_with_cse} shows the effect of scaling the cost of communication between the sensing devices and cloud inference service, $C_{S_iE}$. In this case, we consider $Y \in\{1,2\},\lambda=0.85,\delta_\mu=1$ for both $N=4$ and $N=8$. Predictably, the average overall cost of our framework increases with increasing $C_{S_iE}$. Importantly, so long as  $C_{S_iE}>C_{S_iS_j}$, the average cost for our framework remains both lower and scales more slowly than the cloud-only baseline. In terms of accuracy, we see that for low values of $C_{S_iE}$, we observe lower accuracy -- with this effect being more pronounced for $N=8$. We theorize that this is possibly caused by a larger number of offloads to the edge server, which neutralises a possible regularising effect caused by the action of local inferences.

\subsection{Effect of Computation Cost}\label{comp_cost_scale}
The effect of incorporating the energy consumption of the computation stage is strongly application dependant -- being determined by factors including the specific classification methods used, details of their implementation, and underlying hardware. However, our framework can be trivially extended to incorporate this consideration, so long as an estimate of computation cost can be determined for each of the three possible actions within our proposed approach -- performing direct inference, requesting data and evaluating, and offloading to the cloud. If so, these can simply be included in our estimates of $C_{S_iS_j}$ and $C_{S_iE}$. Based on our observations in Fig.~\ref{fig:scaling_with_cse}, so long as $C_{S_iE}>C_{S_iS_j}$ holds for our revised estimates, our framework will be useful and scalably outperform a traditional centralised inference approach.

\subsection{Effect of Target Device $T$}                                                                                                                                                    
Similarly to the effect of computation cost in \ref{comp_cost_scale}, the effect of the target device $T$ on our framework is determined by incorporating the estimated cost of transferring data from sensors to $T$ into \ref{req_eqn}, which becomes:

\begin{equation}
    \left(C_{S_iE}+C_{S_iT}\right)\,\hat{p}_{ij}+(C_{s_is_j}+C_{S_iE})(1-\hat{p_{ij}})\leq C_{s_iE}
\end{equation}

We again expect that, at worst, if $C_{S_iE}>C_{S_iS_j}+C_{S_iT}$, our framework will still scale usefully compared to the cloud baseline. In practice, this corresponds to a situation where the target device exists with a lower cost cross-link to the sensor devices -- such as the case of a wireless sensing network where it is situated in the same geographic region, and accessible via LoRa. In practice, for simple inference tasks such as the Gaussian cases we have discussed in this paper, it is not meaningful to assume such a target device -- if present, this device could act as a centralised low-cost compute service that negates the need for a high cost uplink to access cloud services. For more complex inference tasks, such as deep learning tasks, such a target device may exist with low communication cost, and be able to aggregate reported inference results/decision, but not have the compute capacity to perform joint inference. In this instance, including a high cost uplink to some cloud or other centralised compute platform is still required, and our framework may have utility.

\subsection{Number of Classes}

\begin{figure}
    \centering
    \includegraphics[width=\linewidth]{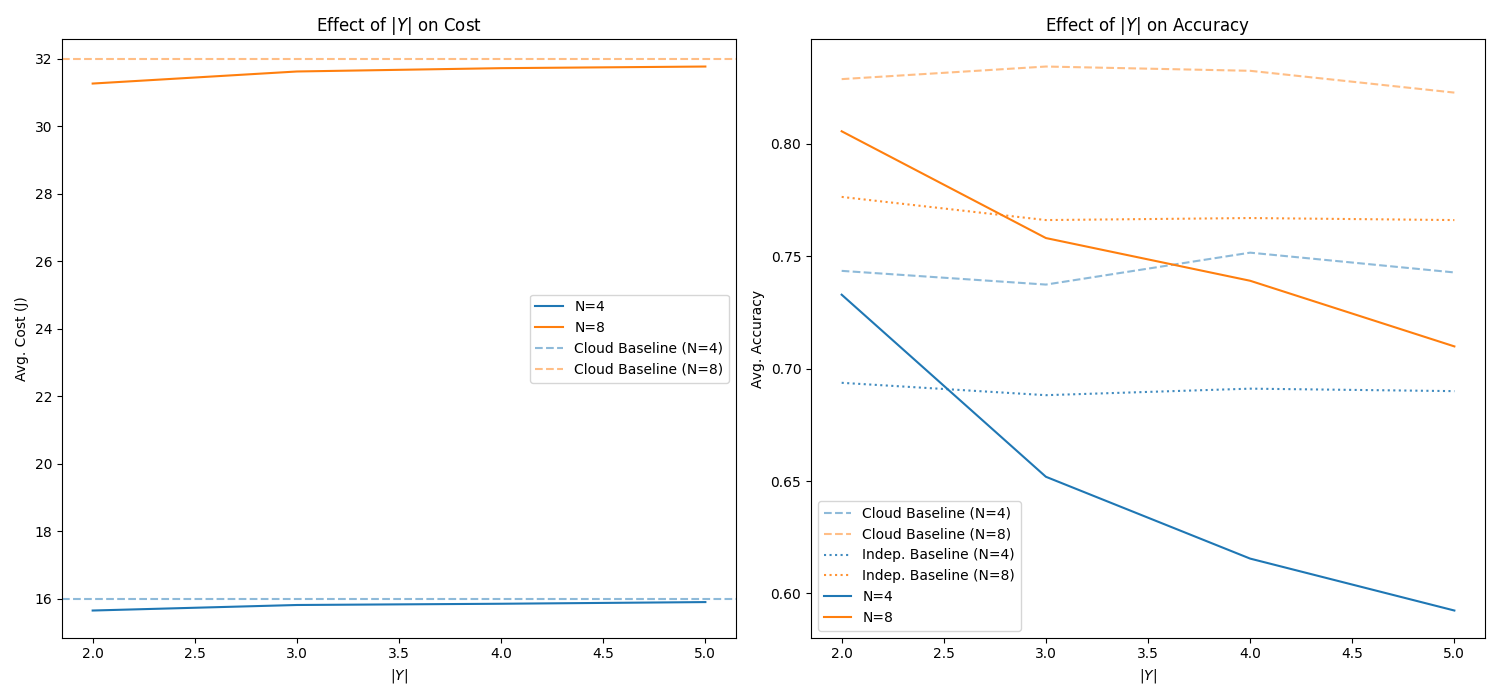}
    \caption{Visualising the effect of increasing the domain of $Y$ on overall cost and accuracy (assuming $C_{S_iS_j}=1J$, $C_{S_iE}=4J$. As expected, we see that this has increasing cost, with decreasing accuracy.}
    \label{fig:scaling_with_classes}
\end{figure}

Fig.~\ref{fig:scaling_with_classes} shows the effect of scaling the domain of $Y$, for both $N=4$ and $N=8$, with $C_{S_iS_j}=1,C_{S_iS_j}=4,\lambda=0.85,\delta_\mu=1$. The cost scales in both cases, as could be expected with increasing complexity of the underlying inference task, but remains below the centralised baseline. Accuracy underperforms both baselines for all $|dom(Y)|>2$. We suspect that $\delta_\mu=1$ does not provide sufficient separation between state distributions for $|dom(Y)|>2$. This indicates that while $\delta_\mu$ largely dictates the effectiveness our our approach, its effect seems to be strongly dependant on the size of the state space.

\section{Conclusions, Recommendations, \& Future Research}\label{conclusions}

\subsection{Conclusions}

We have proposed and defined a hybrid framework for collective inference over networked sensors that incorporates cloud and edge computing, as well as distributed inference. Under assumptions of Gaussian sensor data, we have comprehensively explored the potential benefits of our proposed framework.

We have defined an exemplar theoretical network context, with a model for inter-device cost of communication.

We have shown, by means of analytical derivations, numerical evaluations and experimental evaluations, that there are certain parameterisations of the underlying data for which our framework is highly effective -- achieving similar levels of predictive accuracy as a centralised cloud/fog approach, at a much lower level of cost.

We have further explored the scalability and robustness of our framework, and determined that it has the potential for strong relevance to large-scale networks of sensors.

\subsection{Future Research}

\subsubsection{Measures of Cost}
The performance of our framework is critically dependent on our cost metric, which is currently very simplistic. For any practical scenario, there are further considerations that may need to be incorporated into a new and improved cost metric:

\begin{enumerate}
    \item More realistic estimates of communication energy. This will likely rely on implementation on physical hardware, or detailed simulation using tools such as NS-3, for a more robust cost metric.
    \item The cost of computation. While we briefly consider the cost of computation, this is a critical component of any real-world application. For Deep Learning methods, the use of accelerators, and emerging low-cost inference techniques in TinyML would have an important role in the overall efficiency of sophisticated collective inference models.
    \item Factors other than energy. Bandwidth, latency, and other metrics of resource utilisation are also important factors in many applications. The benefit of our framework for the optimization of these metrics seems intuitive, but requires further investigation.
\end{enumerate}

\subsubsection{More Complex Classification Methods}
In this paper, we have considered solely Gaussian sensor data. While our specific implementation of our proposed framework may be directly applicable to a limited number of applications, we expect that a general form of this framework will be mainly useful only when applied with more sophisticated classifiers, determining estimates of
 $P(Y|s_i),\hat{p}_{ij},P(Y|s_i,s_j)$, and $P(Y|\cup_{k\subset N} \{s_k\})$. 

 As a direct extension of our approach, we hypothesise that our heuristic defined in (\ref{heur_algo}) may be extended to arbitrary sensor distributions, down to approximation of the data using Gaussian mixture models. Under this approach, we expect that we would need to run our heuristic over each mode of the GMMs, resulting in computation cost that scales linearly (within each sensor) in the number of mixing coefficients. Whether this approach is feasible or efficient will be determined by future research.
 
\subsubsection{Validation on a Real World Application}
While theorized, and comprehensively evaluated on simulated Gaussian data, it remains for future work to implement this framework on a real-world scenario -- processing real-world data, and communicating over real-world communication channels. Future research that demonstrates that compares the performance of our framework in this context against existing state-of-the-art approaches will provide insights into how well it can actualise the potential that we have demonstrated here.

\FloatBarrier

\printbibliography

\end{document}